\documentclass[twocolumn,showpacs,prd,preprintnumbers,superscriptaddress]{revtex4-1}

\usepackage{graphicx}
\usepackage{epsfig}
\usepackage{xspace}
\usepackage{dcolumn}
\usepackage{multirow}
\usepackage{longtable}
\usepackage{amsmath}
\usepackage{adjustbox}
\usepackage{tikz}
\usepackage{lineno}
\usepackage{textgreek}
\usetikzlibrary{er,positioning}
\usepackage{float}

\newcolumntype{d}[1]{D{.}{\cdot}{#1}}
\newcolumntype{.}{D{.}{.}{-1}}
\newcolumntype{,}{D{,}{,}{2}}

\begin{document}

\title{A Measurement of the Tau Neutrino Cross Section in Atmospheric Neutrino Oscillations with Super-Kamiokande}

\newcommand{\AFFicrr}{\affiliation{Kamioka Observatory, Institute for Cosmic Ray Research, University of Tokyo, Kamioka, Gifu 506-1205, Japan}}
\newcommand{\AFFkashiwa}{\affiliation{Research Center for Cosmic Neutrinos, Institute for Cosmic Ray Research, University of Tokyo, Kashiwa, Chiba 277-8582, Japan}}
\newcommand{\AFFipmu}{\affiliation{Kavli Institute for the Physics and
Mathematics of the Universe (WPI), The University of Tokyo Institutes for Advanced Study,
University of Tokyo, Kashiwa, Chiba 277-8583, Japan }}
\newcommand{\AFFmad}{\affiliation{Department of Theoretical Physics, University Autonoma Madrid, 28049 Madrid, Spain}}
\newcommand{\AFFubc}{\affiliation{Department of Physics and Astronomy, University of British Columbia, Vancouver, BC, V6T1Z4, Canada}}
\newcommand{\AFFbu}{\affiliation{Department of Physics, Boston University, Boston, MA 02215, USA}}
\newcommand{\AFFuci}{\affiliation{Department of Physics and Astronomy, University of California, Irvine, Irvine, CA 92697-4575, USA }}
\newcommand{\AFFcsu}{\affiliation{Department of Physics, California State University, Dominguez Hills, Carson, CA 90747, USA}}
\newcommand{\AFFcnm}{\affiliation{Department of Physics, Chonnam National University, Kwangju 500-757, Korea}}
\newcommand{\AFFduke}{\affiliation{Department of Physics, Duke University, Durham NC 27708, USA}}
\newcommand{\AFFfukuoka}{\affiliation{Junior College, Fukuoka Institute of Technology, Fukuoka, Fukuoka 811-0295, Japan}}
\newcommand{\AFFgifu}{\affiliation{Department of Physics, Gifu University, Gifu, Gifu 501-1193, Japan}}
\newcommand{\AFFgist}{\affiliation{GIST College, Gwangju Institute of Science and Technology, Gwangju 500-712, Korea}}
\newcommand{\AFFuh}{\affiliation{Department of Physics and Astronomy, University of Hawaii, Honolulu, HI 96822, USA}}
\newcommand{\AFFicl}{\affiliation{Department of Physics, Imperial College London , London, SW7 2AZ, United Kingdom }}
\newcommand{\AFFkek}{\affiliation{High Energy Accelerator Research Organization (KEK), Tsukuba, Ibaraki 305-0801, Japan }}
\newcommand{\AFFkobe}{\affiliation{Department of Physics, Kobe University, Kobe, Hyogo 657-8501, Japan}}
\newcommand{\AFFkyoto}{\affiliation{Department of Physics, Kyoto University, Kyoto, Kyoto 606-8502, Japan}}
\newcommand{\AFFliv}{\affiliation{Department of Physics, University of Liverpool, Liverpool, L69 7ZE, United Kingdom}}
\newcommand{\AFFmiyagi}{\affiliation{Department of Physics, Miyagi University of Education, Sendai, Miyagi 980-0845, Japan}}
\newcommand{\AFFnagoya}{\affiliation{Institute for Space-Earth Enviromental Research, Nagoya University, Nagoya, Aichi 464-8602, Japan}}
\newcommand{\AFFkmi}{\affiliation{Kobayashi-Maskawa Institute for the Origin of Particles and the Universe, Nagoya University, Nagoya, Aichi 464-8602, Japan}}
\newcommand{\AFFpol}{\affiliation{National Centre For Nuclear Research, 00-681 Warsaw, Poland}}
\newcommand{\AFFsuny}{\affiliation{Department of Physics and Astronomy, State University of New York at Stony Brook, NY 11794-3800, USA}}
\newcommand{\AFFokayama}{\affiliation{Department of Physics, Okayama University, Okayama, Okayama 700-8530, Japan }}
\newcommand{\AFFosaka}{\affiliation{Department of Physics, Osaka University, Toyonaka, Osaka 560-0043, Japan}}
\newcommand{\AFFox}{\affiliation{Department of Physics, Oxford University, Oxford, OX1 3PU, United Kingdom}}
\newcommand{\AFFqmul}{\affiliation{School of Physics and Astronomy, Queen Mary University of London, London, E1 4NS, United Kingdom}}
\newcommand{\AFFregina}{\affiliation{Department of Physics, University of Regina, 3737 Wascana Parkway, Regina, SK, S4SOA2, Canada}}
\newcommand{\AFFseoul}{\affiliation{Department of Physics, Seoul National University, Seoul 151-742, Korea}}
\newcommand{\AFFsheff}{\affiliation{Department of Physics and Astronomy, University of Sheffield, S10 2TN, Sheffield, United Kingdom}}
\newcommand{\AFFshizuokasc}{\affiliation{Department of Informatics in
Social Welfare, Shizuoka University of Welfare, Yaizu, Shizuoka, 425-8611, Japan}}
\newcommand{\AFFstfc}{\affiliation{STFC, Rutherford Appleton Laboratory, Harwell Oxford, and Daresbury Laboratory, Warrington, OX11 0QX, United Kingdom}}
\newcommand{\AFFskk}{\affiliation{Department of Physics, Sungkyunkwan University, Suwon 440-746, Korea}}
\newcommand{\AFFtokyo}{\affiliation{The University of Tokyo, Bunkyo, Tokyo 113-0033, Japan }}
\newcommand{\AFFtodai}{\affiliation{Department of Physics, University of Tokyo, Bunkyo, Tokyo 113-0033, Japan }}
\newcommand{\AFFtit}{\affiliation{Department of Physics,Tokyo Institute of Technology, Meguro, Tokyo 152-8551, Japan }}
\newcommand{\AFFtus}{\affiliation{Department of Physics, Faculty of Science and Technology, Tokyo University of Science, Noda, Chiba 278-8510, Japan }}
\newcommand{\AFFtoronto}{\affiliation{Department of Physics, University of Toronto, ON, M5S 1A7, Canada }}
\newcommand{\AFFtriumf}{\affiliation{TRIUMF, 4004 Wesbrook Mall, Vancouver, BC, V6T2A3, Canada }}
\newcommand{\AFFtokai}{\affiliation{Department of Physics, Tokai University, Hiratsuka, Kanagawa 259-1292, Japan}}
\newcommand{\AFFtsinghua}{\affiliation{Department of Engineering Physics, Tsinghua University, Beijing, 100084, China}}
\newcommand{\AFFynu}{\affiliation{Faculty of Engineering, Yokohama National University, Yokohama, 240-8501, Japan}}
\newcommand{\AFFuw}{\affiliation{Department of Physics, University of Washington, Seattle, WA 98195-1560, USA}}

\AFFicrr
\AFFkashiwa
\AFFmad
\AFFbu
\AFFubc
\AFFuci
\AFFcsu
\AFFcnm
\AFFduke
\AFFfukuoka
\AFFgifu
\AFFgist
\AFFuh
\AFFicl
\AFFkek
\AFFkobe
\AFFkyoto
\AFFliv
\AFFmiyagi
\AFFnagoya
\AFFkmi
\AFFpol
\AFFsuny
\AFFokayama
\AFFosaka
\AFFox
\AFFqmul
\AFFregina
\AFFseoul
\AFFsheff
\AFFshizuokasc
\AFFstfc
\AFFskk
\AFFtokai
\AFFtokyo
\AFFtodai
\AFFipmu
\AFFtit
\AFFtus
\AFFtoronto
\AFFtriumf
\AFFtsinghua
\AFFynu

\author{Z.~Li}
\AFFduke
\author{K.~Abe}
\AFFicrr
\AFFipmu
\author{C.~Bronner}
\AFFicrr
\author{Y.~Hayato}
\AFFicrr
\AFFipmu
\author{M.~Ikeda}
\AFFicrr
\author{K.~Iyogi}
\AFFicrr 
\author{J.~Kameda}
\AFFicrr
\AFFipmu 
\author{Y.~Kato}
\AFFicrr
\author{Y.~Kishimoto}
\AFFicrr
\AFFipmu 
\author{Ll.~Marti}
\AFFicrr
% ...

\collaboration{The Super-Kamiokande Collaboration}
\noaffiliation

\date{\today}

\begin{abstract}
Using 5,326 days of atmospheric neutrino data, a search for atmospheric tau neutrino appearance has been performed in the Super-Kamiokande experiment.  Super-Kamiokande measures the tau normalization to be 1.47$\pm$0.32 under the assumption of normal neutrino hierarchy, relative to the expectation of unity with neutrino oscillation. The result excludes the hypothesis of no-tau-appearance with a significance level of 4.6$\sigma$. The inclusive charged-current tau neutrino cross section averaged by the tau neutrino flux at Super-Kamiokande is measured to be $(0.94\pm0.20)\times 10^{-38}$ cm$^{2}$. The measurement is consistent with the Standard Model prediction, agreeing to within 1.5$\sigma$.

\end{abstract}


\maketitle

\section{Introduction}\label{sec:intro}
\par In the three-flavor neutrino framework, the three neutrino flavor states ($
\nu_e$, $\nu_\mu$, $\nu_\tau$) are superpositions of three neutrino mass states ($\nu_1$, $\nu_2$, $\nu_3$). The oscillation parameters in the framework have been measured in atmospheric neutrino experiments \citep{Fukuda:1998mi, Abe:2011ph, Aartsen:2016psd},  solar neutrino experiments \cite{Ahmad:2001an,Fukuda:2001nj,Ahmad:2002jz}, reactor neutrino experiments \cite{Eguchi:2002dm,An:2012eh,Abe:2011fz,Ahn:2012nd}, and long-baseline neutrino experiments \cite{Ahn:2002up,Michael:2006rx,Abe:2012gx}. In the three-flavor neutrino framework, the deficit of atmospheric muon neutrinos observed in the Super-Kamiokande experiment \citep{Fukuda:1998mi} can be explained by the change of muon neutrinos to tau neutrinos during their propagation. A direct detection of tau neutrinos from neutrino oscillation is important for an unambiguous confirmation of three-flavor neutrino oscillations. 
\par However, the detection of tau neutrino appearance is challenging. Charged-current neutrino interactions are required to determine the flavor in neutrino detection. Charged-current tau lepton appearance has an energy threshold of 3.5 GeV, and the charged-current tau neutrino cross section is greatly suppressed at low energies due to the large mass of the tau lepton relative to the electron or muon. The DONUT experiment first directly observed the tau neutrino by measuring charged-current interactions using a high-energy neutrino beam that contained tau neutrinos \citep{Kodama:2000mp}. Long-baseline experiments tuned for maximum oscillation have the bulk of their neutrinos below this energy. In addition, the tau lepton has an extremely short lifetime, making a direct detection very difficult. Nevertheless, the long-baseline neutrino experiment OPERA measured tau neutrino appearance in a high-energy muon neutrino beam by observing 5 $\nu_\tau$ events with a background expectation of 0.25 events \cite{Agafonova:2015jxn}. 
\par Atmospheric neutrinos are mostly electron or muon neutrinos at production \cite{Honda:2015fha}. Tau neutrino appearance is expected in the atmospheric neutrinos from neutrino oscillations. In three-flavor neutrino oscillation in the vacuum, the probability of $\nu_\tau$ appearance can be approximately expressed as%
\begin{subequations}
\begin{equation}
P_{\nu_{\mu}\rightarrow\nu_{\tau}}\simeq\cos^2{\theta_{13}}\sin^2(2\theta_{23})\sin^2(1.27\Delta m^2_{32} \frac{L}{E}),
\end{equation}
\begin{equation}
P_{\nu_{e}\rightarrow\nu_{\tau}}\simeq\sin^2(2\theta_{13})\cos^2(\theta_{23})\sin^2(1.27\Delta m^2_{32}\frac{L}{E}),
\end{equation}
\label{eq:tauprob}
\end{subequations}%

\noindent where $\Delta m_{32}^2\equiv m^2_3-m^2_2$ is the mass splitting in eV$^2$, $\theta_{ij}$ is a mixing angle in the PMNS matrix, $L$ is neutrino path length in km, and $E$ is neutrino energy in GeV. Atmospheric neutrinos have energies spanning many orders of magnitude from 10 MeV to more than 1 TeV; the high energy component of the atmospheric neutrinos have enough energy for charged-current tau neutrino interactions. Super-Kamiokande is expected to detect roughly one charged-current tau neutrino interaction per kiloton of water per year. Super-Kamiokande previously published a measurement of  atmospheric tau neutrino appearance consistent with three-flavor neutrino oscillation with data collected in SK-I through SK-III
\cite{Abe:2012jj}. This analysis has been updated with data collected in SK-IV between 2008 and 2016, and the simulation and reconstruction have been improved. Using the measured charged-current tau neutrino events, Super-Kamiokande also measures the charged-current tau neutrino cross section. 
\par This paper proceeds as follows: Section \ref{sec:superk} describes some basic features of the Super-Kamiokande experiment (Super-K, SK). Section \ref{sec:mc} describes Monte Carlo simulations of both the charge-current tau neutrino signal and the atmospheric neutrino background. Section \ref{sec:reducrecon} describes standard data selection and reconstruction algorithms used in the analysis. Section \ref{sec:taunn} describes a neural network algorithm developed to select the tau
signal. Section \ref{sec:analysis} describes a search for atmospheric tau neutrino appearance, and a measurement of charge-current tau neutrino cross section. Section \ref{sec:conclusion} presents our results and conclusion.

\section{The Super-Kamiokande Detector}\label{sec:superk}
Super-Kamiokande is a 50 kiloton cylindrical water-Cherenkov detector located in the Kamioka mine under about 1 km rock overburden (2.7 km water equivalent) at the Ikenoyama mountain in Japan \citep{Fukuda:2002uc, Abe:2013gga}. The detector is arranged into two optically-separated regions: the inner detector (ID) and the outer detector (OD). The ID is instrumented with 11,129 20-inch inward-facing PMT's and the OD is instrumented with 1,885 8-inch outward-facing PMT's. The PMTs collect Cherenkov light produced in the ultra-pure water in the detector. A fiducial volume of the ID is
defined as the cylindrical volume 2 meter inward from the ID wall, and has a mass of 22.5 kilotons\citep{Fukuda:2002uc}.
\par Super-K has been in operation since 1996, and has had four data-taking periods. The first period, called SK-I, began in April 1996, with 11,146 PMT's covering 40$\%$ of the ID surface. The SK-I period continued until July 2001, totaling 1489.2 live-days. An accident in November 2001 destroyed half of the ID PMT's. The remaining 5,182 PMT's were rearranged uniformly on the ID surface, covering 19$\%$ of the surface. The data-taking period with this decreased photo-coverage between
December 2002 and October 2005 is called SK-II. This period lasted 798.6 live-days. A full reconstruction of the detector restored the photo-coverage to 40$\%$ in 2006. The third data-taking period, called SK-III, lasted between July 2006 and September 2008, comprising 518.1 live-days. The detector was upgraded with improved electronics in the summer of 2008 \cite{Yamada:2010zzc}. The period after the upgrade is referred as SK-IV. In this paper, SK-IV data are used up to March
2016, totaling 2519.9 live-days. The complete SK-I through SK-IV data set comprises a total exposure of 5,326 live-days. 

\section{Simulation}\label{sec:mc}
In order to predict the rate of both tau signal and atmospheric neutrino background, a full Monte Carlo (MC) simulation is used to model both the neutrino interactions and the detector response of Super-K. Since the four Super-K periods have different detector configurations, separate sets of MC for both tau signal and atmospheric neutrino background are generated for each period.

\par Atmospheric neutrinos are produced from the decays of charged mesons and muons in the cosmic-ray induced atmospheric showers, and are mostly $\nu_\mu$ and $\nu_e$ at production. The intrinsic tau neutrinos in the atmospheric neutrino flux are negligible for this analysis\citep{Martin:2003us}. Three-dimensional neutrino fluxes of $\nu_\mu$ and $\nu_e$ are modeled from the calculation of Honda $et$ $al.$ \cite{Honda:2015fha}. The calculation predicts the fluxes of electron and muon neutrinos as a function of neutrino direction and neutrino energy at the Super-K site. 

\subsection{Neutrino fluxes}
\par
\begin{figure}[!ht]
\centering
\includegraphics[width=0.4\textwidth]{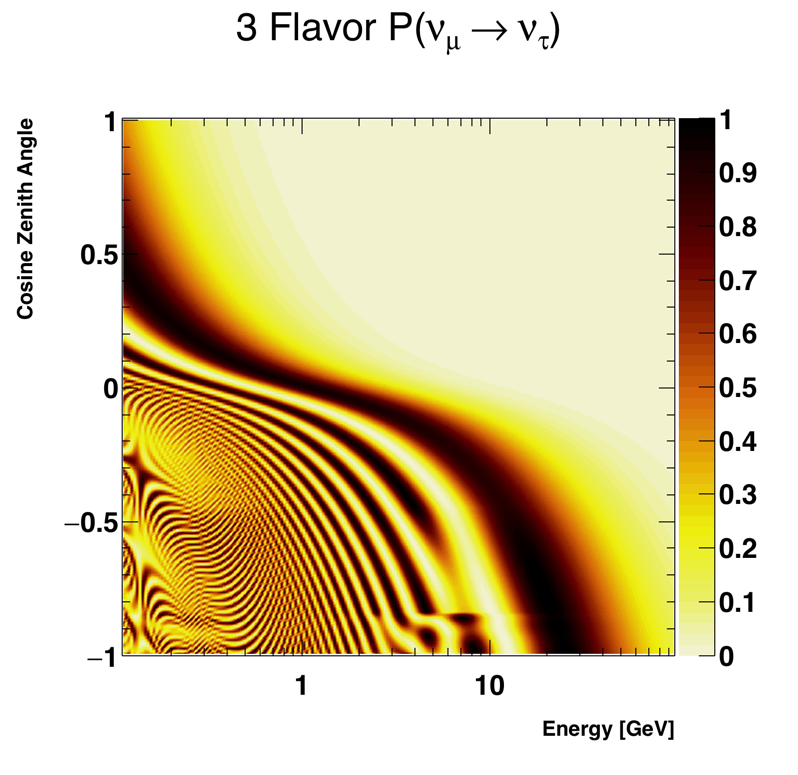}
~
\centering
\includegraphics[width=0.4\textwidth]{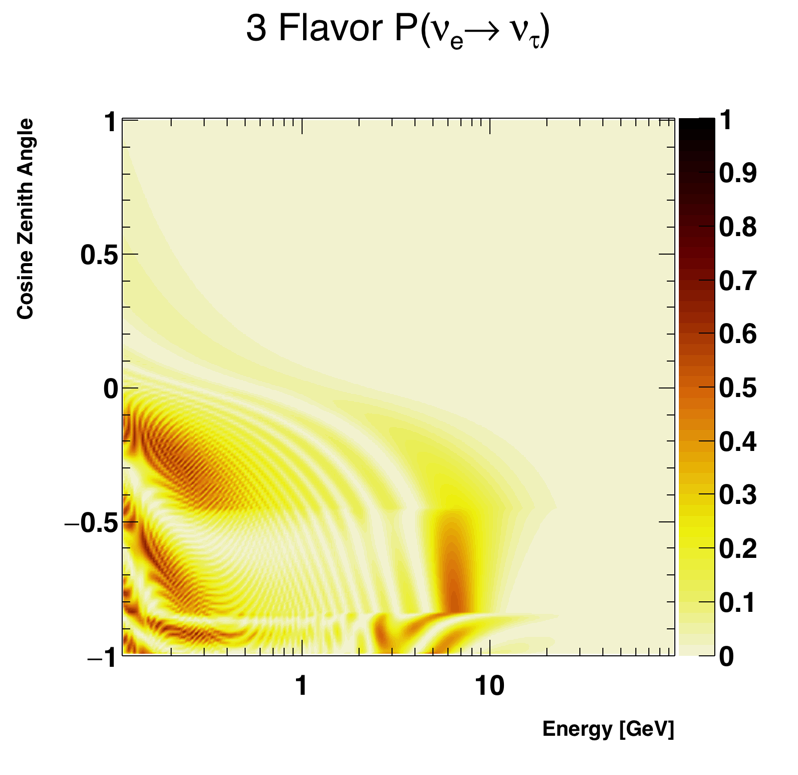}

\caption{Probabilities of tau neutrino appearance from muon neutrino (top) or electron neutrino (bottom) as a function of neutrino energy and zenith angle, $\Delta m_{32}^2=2.1\times 10^{-3} $eV$^2$, $\Delta m_{21}^2=7.6\times10^{-5} $eV$^2$, $\sin^22\theta_{23}=1.0$, $\sin^22\theta_{13}=0.099$, $\delta_{CP}=0$ and assuming the normal hierarchy. Cosine of zenith angle equal to 1 corresponds to downward-going direction of neutrinos, and cosine of zenith angle equal to -1 corresponds to upward-going direction of neutrinos. }
\label{fig:tauprob}
\end{figure}

\begin{figure}[!ht]
\centering
\includegraphics[width=0.4\textwidth]{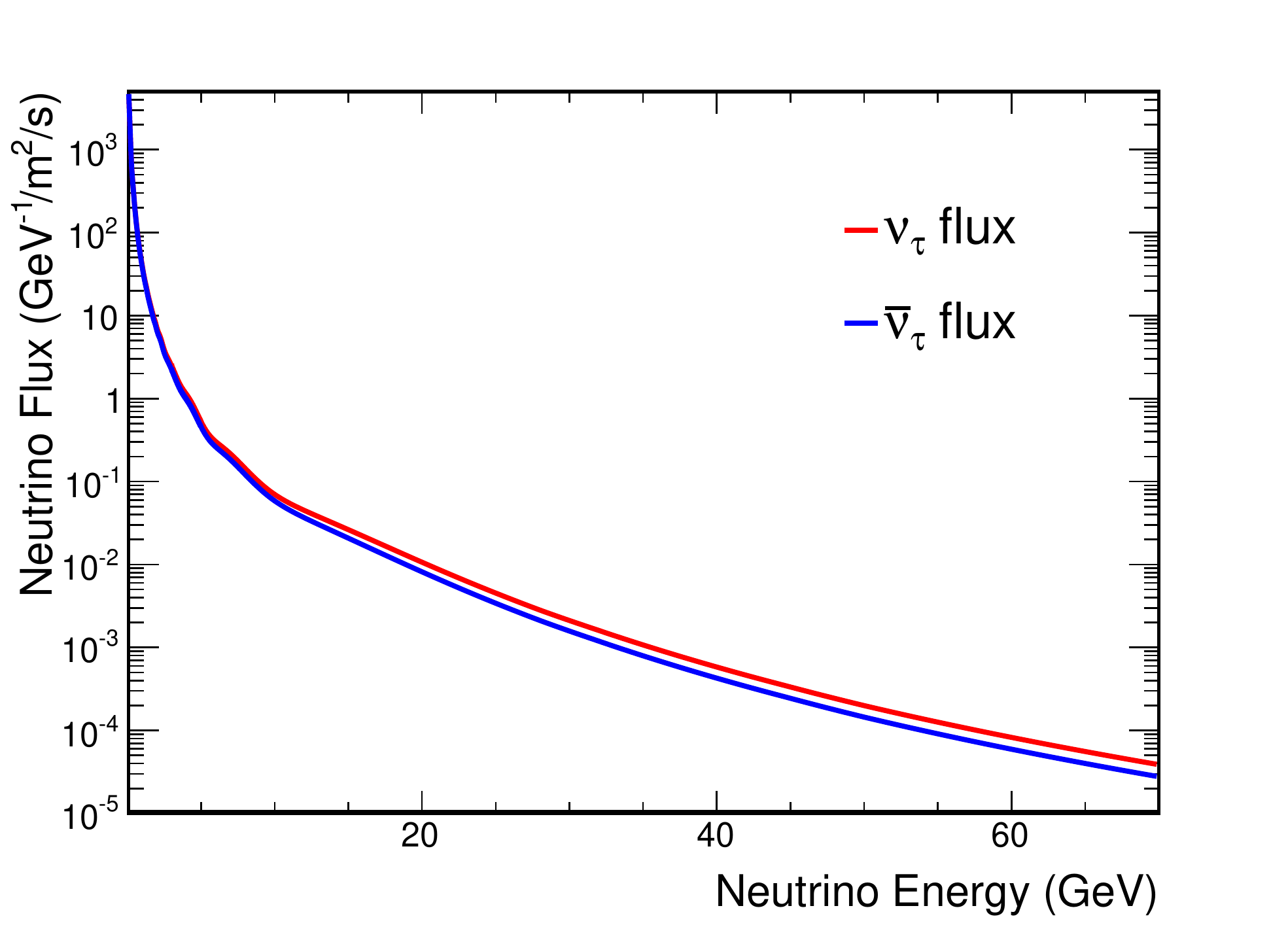}
\includegraphics[width=0.4\textwidth]{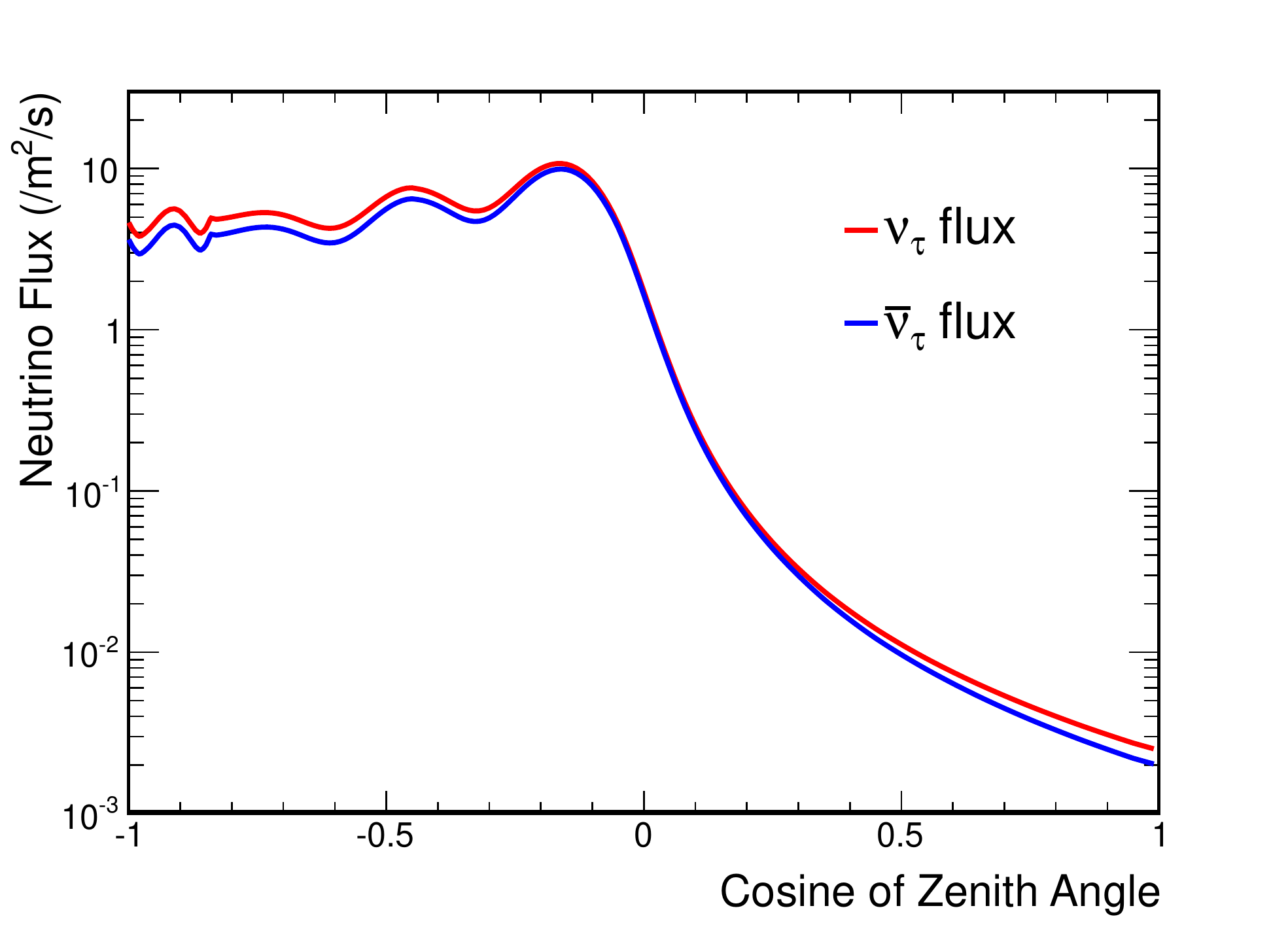}
\caption{Fluxes of atmospheric $\nu_\tau$ (red) and $\bar\nu_\tau$ (blue) from neutrino oscillations as a function of neutrino energy (top) and cosine of zenith angle (bottom) at Super-Kamiokande based on Honda flux calculation \citep{Honda:2015fha} and tau neutrino appearance probabilities.}
\label{fig:taufluxosc}
\end{figure}

 Although atmospheric neutrinos consist of $\nu_\mu$ and $\nu_e$ at production, $\nu_\tau$ are expected to appear due to neutrino oscillations. The probabilities of $\nu_\tau$ appearance from neutrino oscillations of $\nu_\mu$ or $\nu_e$ in vacuum is shown in Equation \ref{eq:tauprob}. However since neutrinos coming from below travel through the Earth, the oscillation probabilities are altered by the matter effect. Therefore, a custom code \cite{Prob3++} is used to calculate the
oscillation probabilities, which takes into account the effect of neutrino types, path lengths, neutrino energies and the matter effect. The oscillation parameters used are $\Delta m_{32}^2=2.1\times 10^{-3} $eV$^2$, $\Delta m_{21}^2=7.6\times10^{-5} $eV$^2$, $\sin^22\theta_{23}=1.0$, $\sin^22\theta_{13}=0.099$, $\delta_{CP}=0$ \cite{Beringer:1900zz}. A method from \cite{Barger:1980tf} is used to account for the matter effect in the calculation of oscillation probabilities based on the matter density structure of
the Earth in Ref. \cite{dziewonski1981preliminary}.
Figure~\ref{fig:tauprob} illustrates the tau neutrino appearance probabilities from muon neutrino or electron neutrino oscillations in three-flavor neutrino oscillation under the assumption of the normal hierarchy. For a neutrino with a given energy and path length, a muon neutrino has a larger probability than a electron neutrino to change flavor to a tau neutrino. Following the oscillation calculation, we can predict the atmospheric tau neutrino flux at Super-K. Figure \ref{fig:taufluxosc} shows the expected atmospheric tau neutrino fluxes  from neutrino oscillations at Super-K. 

\subsection{Neutrino interactions}
\par
\begin{figure}[!ht]
\centering
\includegraphics[width=0.4\textwidth]{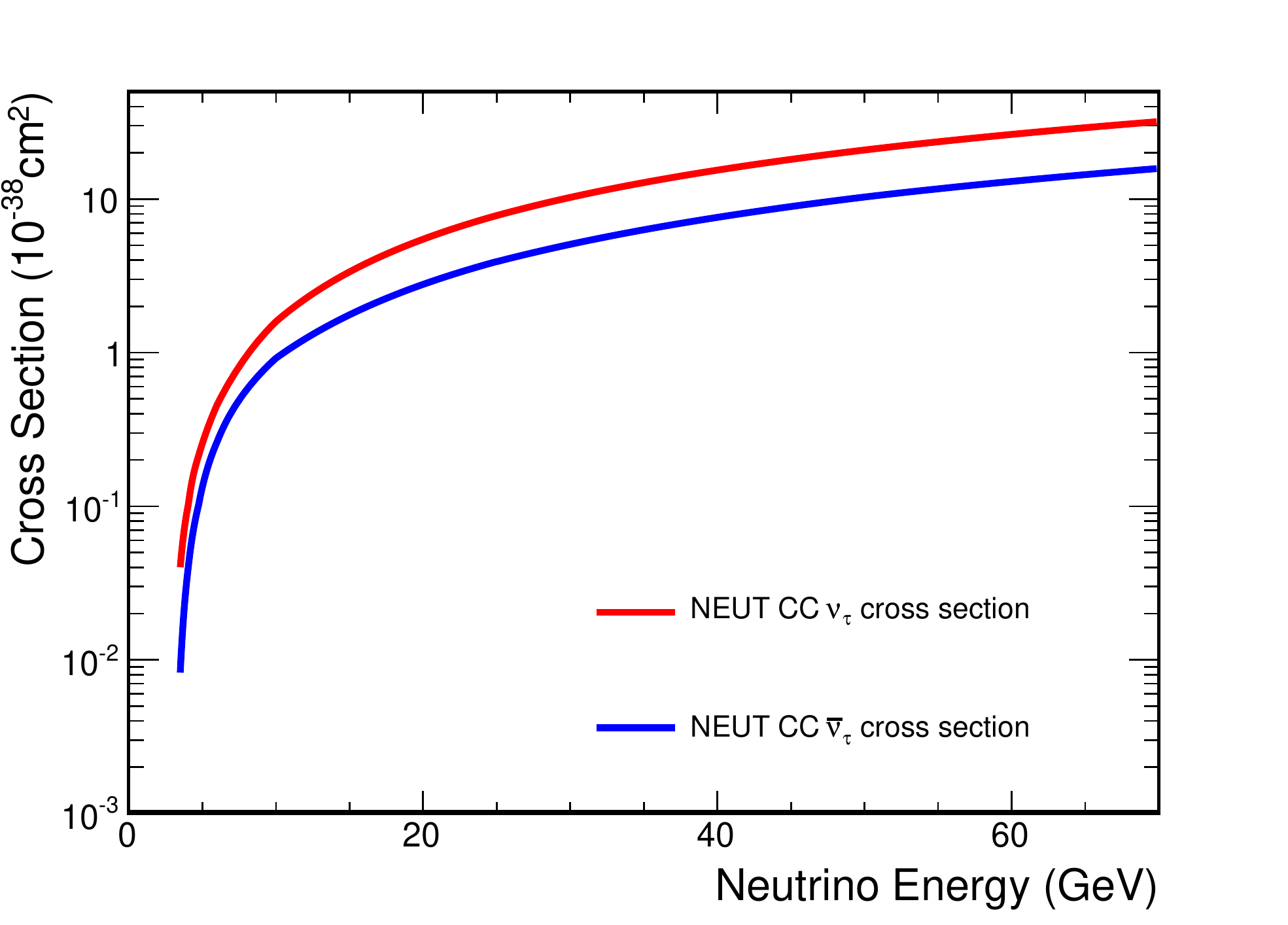}
\caption{Total charged-current cross section as a function of neutrino energy for $\nu_{\tau}$ (red) and $\bar{\nu}_{\tau}$ (blue) from 3.5 GeV to 70 GeV.}
\label{fig:neuttauxsec}
\end{figure}

\begin{figure}[!htbp] 
   \centering
   \includegraphics[width=0.45\textwidth]{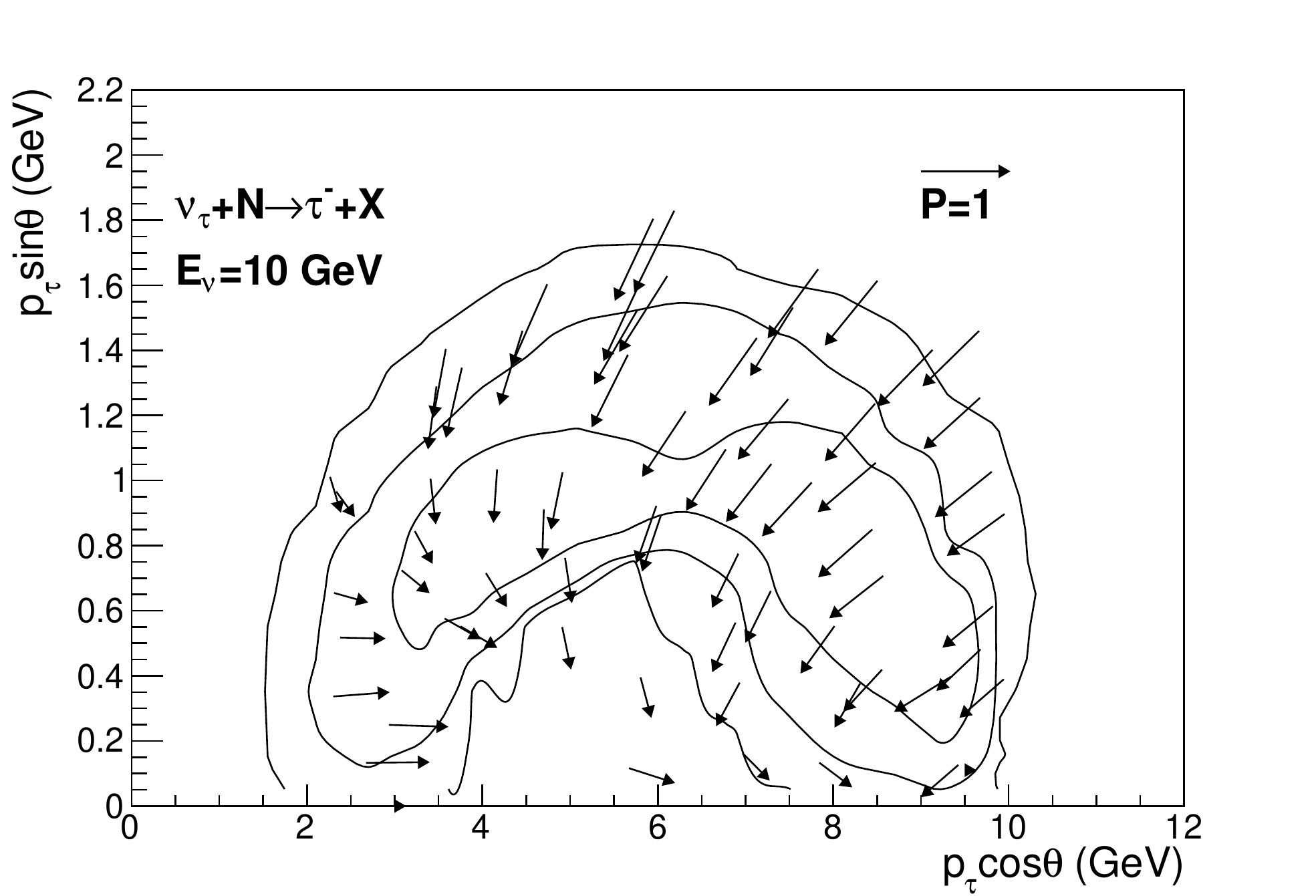} 
   \includegraphics[width=0.45\textwidth]{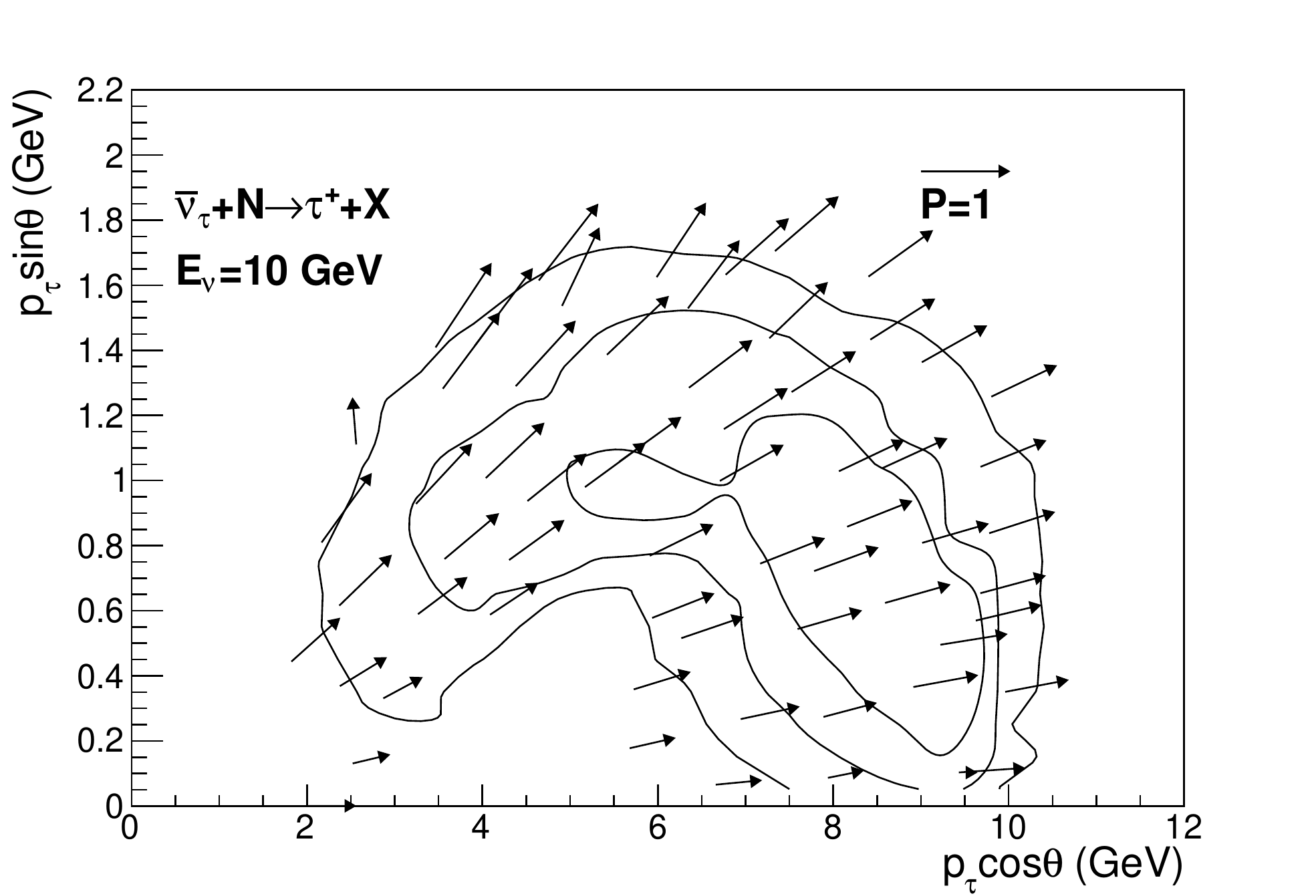} 
   \caption{Polarization of $\tau^-$ (top) and $\tau^+$ (bottom) produced in the interactions of neutrinos with energy of 10 GeV in Super-K simulation. \straighttheta\,is the scattering angle of the tau lepton relative to the neutrino direction. The $\tau$ polarizations are shown by the arrows. The length of the arrows give the degree of polarization, and the direction of arrows give that of the $\tau$ spin in the $\tau$ rest frame. The size of the 100$\%$ polarization ($P=1$) arrow is shown as a reference. The three contours represent regions of 68.3$\%$, 95.4$\%$, and 99.7$\%$ of MC events. }
   \label{fig:taupol}
\end{figure}

 The NEUT code \cite{Hayato:2009zz} is used to model the neutrino nucleon interactions including quasielastic scattering, single meson production, coherent pion production, and deep-inelastic scattering (DIS). In the simulation of atmospheric neutrino background, all $\nu_\mu$ and $\nu_e$ interactions are included. All flavors of neutrinos interact with neutral-current (NC) interactions, and hence are unaffected by oscillations. Atmospheric neutrino eutral current events are simulated based on the total neutrino flux. The simulation of tau signal contains only charged-current (CC) $\nu_\tau$ interactions whose cross sections are calculated following the same models as those used for $\nu_\mu$ and $\nu_e$.  The relatively large mass of the tau lepton produced in the interactions greatly suppresses the cross section of charged-current tau neutrino interactions at low energies and results in an energy threshold of 3.5 GeV. Figure \ref{fig:neuttauxsec} shows the total cross section of charged-current interactions for $\nu_\tau$ and $\bar\nu_\tau$ in the simulations. Tau leptons produced in the CC tau neutrino interactions are polarized, and the polarization affects the distributions of its decay particles. Therefore, a polarization model from Ref. \cite{Hagiwara:2003di} is also included in the simulation. Figure \ref{fig:taupol} shows the polarization of $\tau^-/\tau^+$ in the simulation for interactions of neutrinos with energy of 10 GeV.

\par This analysis selects events at relatively high neutrino energies, at which the CC interactions contains a high percentage of DIS (45$\%$) in the background, with CC $\nu_\tau$ events containing 60$\%$ DIS. The GRV98 \cite{Gluck:1998xa} parton distribution functions are used in the calculation of the DIS cross sections. In order to smoothly match the DIS cross sections with the resonance region, an additional correction developed by Bodek and Yang \cite{Bodek:2002vp} is also applied.

\par The tau lepton has a mean lifetime of 2.9$\times$10$^{-13}$ s and it decays very quickly after production in the detector. The decays of tau lepton are divided into leptonic decays and hadronic decays based on the particles produced. The kinematics of of tau lepton decay is simulated with TAUOLA version 2.6 \cite{Was:2004dg}.
\par The particles produced in both atmospheric neutrino background interactions and CC $\nu_\tau$ interactions are input to a custom detector simulation based on Geant3 \cite{Brun:1987ma}. The code simulates the propagation and Cherenkov light emission of the particles and the Super-K detector \citep{Abe:2013gga}.

\section{Reduction and Reconstruction}\label{sec:reducrecon}
\begin{figure}[htbp] 
   \centering
   \includegraphics[width=0.4\textwidth]{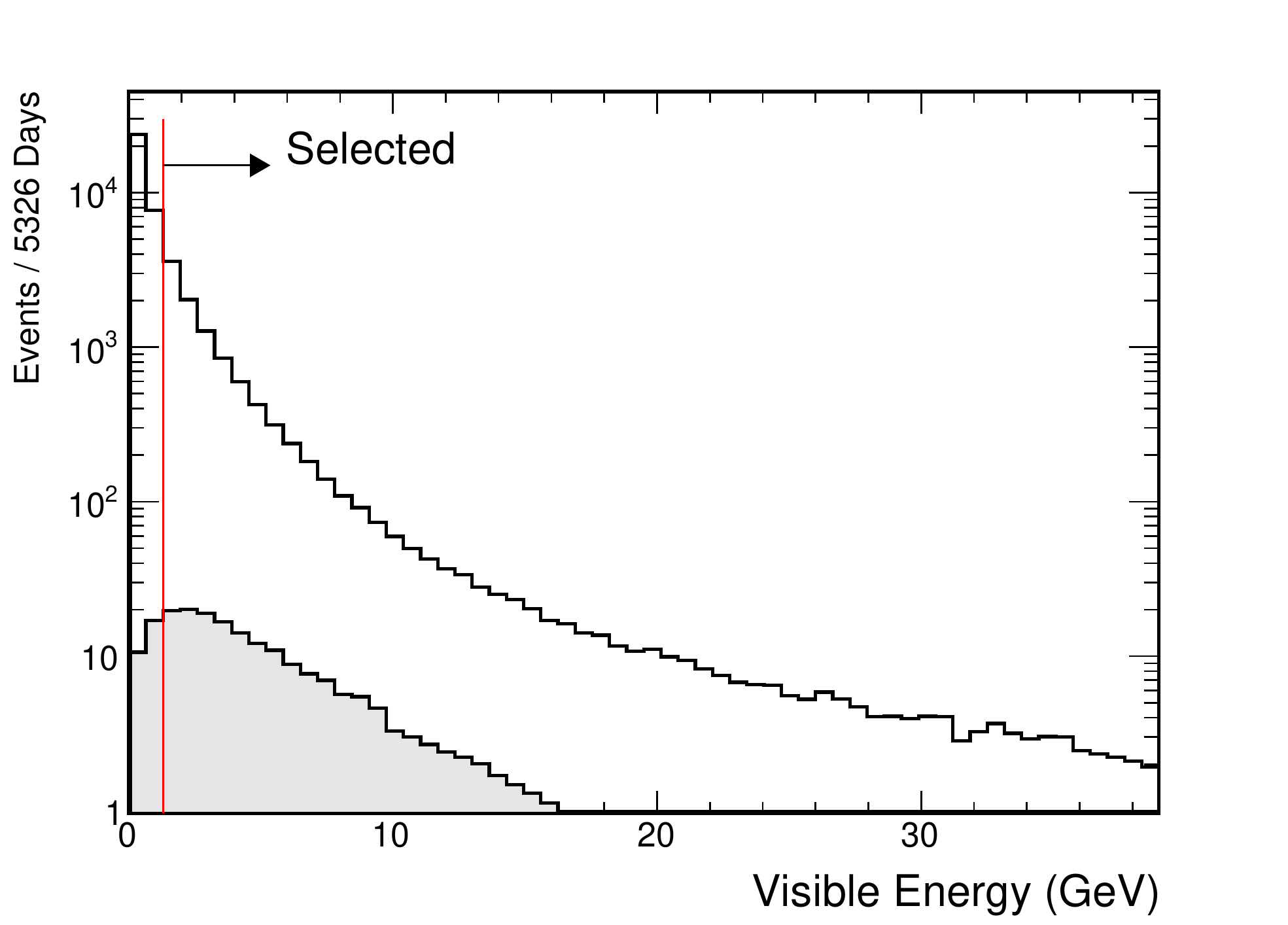} 
   \caption{The distribution of $E_{\rm vis}$ in simulations of atmospheric neutrino background (black histogram) and tau signal (gray shaded histogram) scaled to the live time of SK-I through SK-IV. The atmospheric neutrino background has a bulk of events with $E_{\rm vis}$ less than 1.3 GeV, but the majority of tau signal have $E_{\rm vis}$ more than 1.3 GeV.}
   \label{fig:eviscut}
\end{figure}

This analysis only uses fully-contained (FC) multi-GeV events in the fiducial volume. Fully-contained events are defined as events which only have activity in the ID, and FC events in the fiducial volume are selected by requiring the reconstructed event vertex be at least 2 meters away from the ID wall. In addition, the events are required to have more than 1.3 GeV of visible energy ($E_{\rm vis}$), which is defined as the energy to produce the observed light in the event if it were produced by a
single electron. As shown in Fig. \ref{fig:eviscut}, the $E_{\rm vis}$ cut selects the majority of the tau signal but rejects the bulk of low-energy atmospheric neutrino background events. The selection efficiencies for this set of cuts are 86$\%$ for the $\nu_\tau$ CC signal and 23$\%$ for the background in four Super-K periods. The same selection is applied to the simulations and the observed data.

\par The selected events are passed through a reconstruction program to determine the event vertex, the number of Cherenkov rings, the particle type and momentum of each ring, and the number of Michel electrons. The same reconstruction algorithms are applied to the MCs and the observed data. Events are assumed to originate from a single vertex, and the reconstruction uses the distribution of observed charge and the PMT timing to find the vertex and the brightest Cherenkov ring. A Hough transformation method \cite{davies2004machine} is used to find additional rings. Each ring candidate is tested using a likelihood method to remove fake rings and determine the final number of rings. A likelihood method based on the ring pattern and ring opening angle is used to identify each ring as $e-$like (showering type from $e^\pm$ or $\gamma$) or $\mu-$like (nonshowering type). Michel electrons from stopping muons are tagged by searching for clusters of hits after the primary event. The time window for such clusters extends to 20 $\mu$s after the primary event. In the SK-I to SK-III periods, there was an impedance mismatch in the electronics which caused signal reflection around 1000 ns after the main event. Therefore, the time period 800--1200 ns after the main event was excluded. The improved SK-IV electronics avoids such signal mismatch, thus no exclusion is required. As a result, the tagging efficiency was improved from 80$\%$ to 96$\%$ for $\mu^+$ decays and 63$\%$ to 83$\%$ for $\mu^-$ decays between SK-I-II-III and SK-IV. More details of the reconstruction can be found in Refs. \cite{Shiozawa:1999sd, Ashie:2005ik}.
\par 

\section{A neural network algorithm for tau neutrino search}\label{sec:taunn}
\begin{figure}[htbp]
\centering
\includegraphics[width=0.4\textwidth]{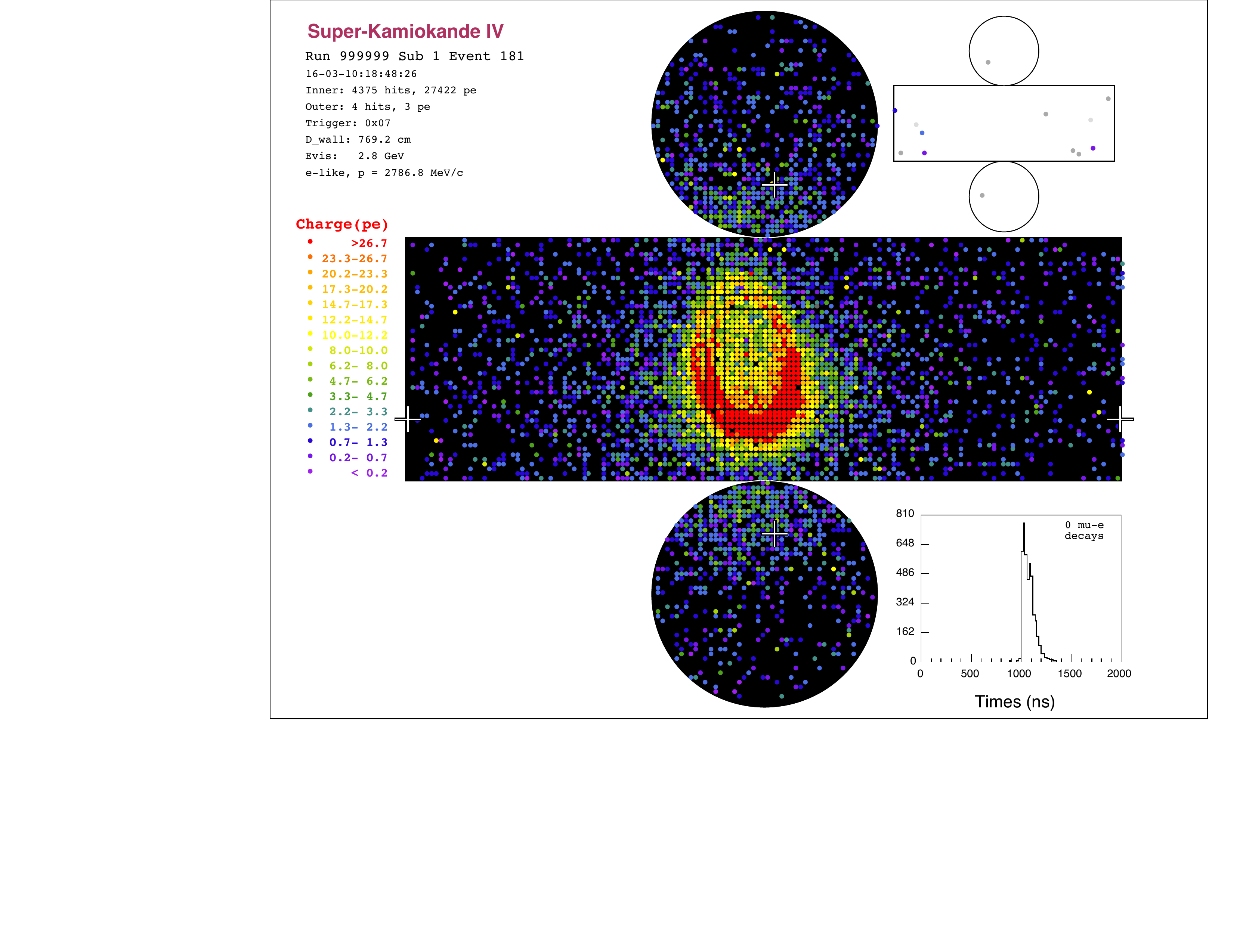}
\includegraphics[width=0.4\textwidth]{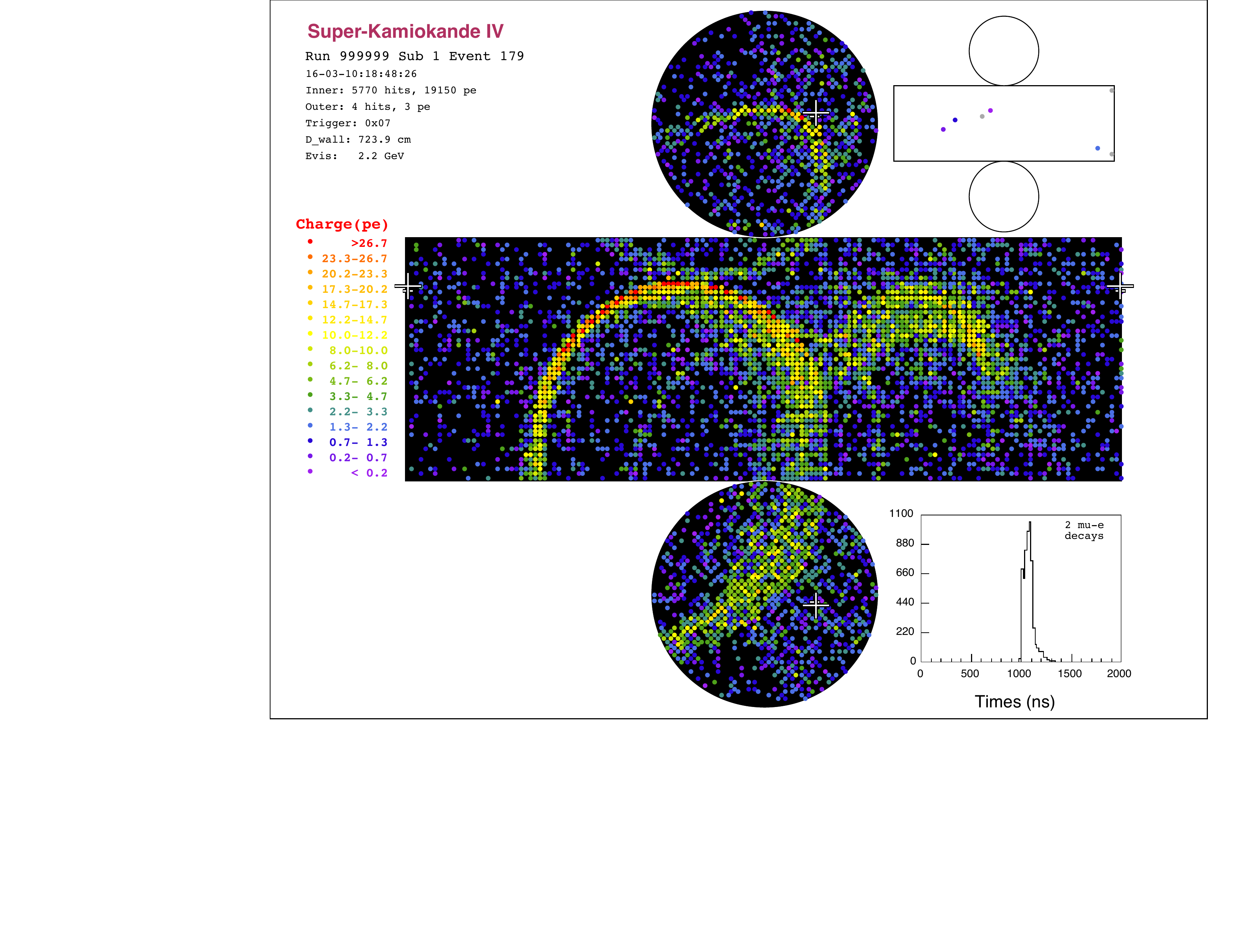}
\includegraphics[width=0.4\textwidth]{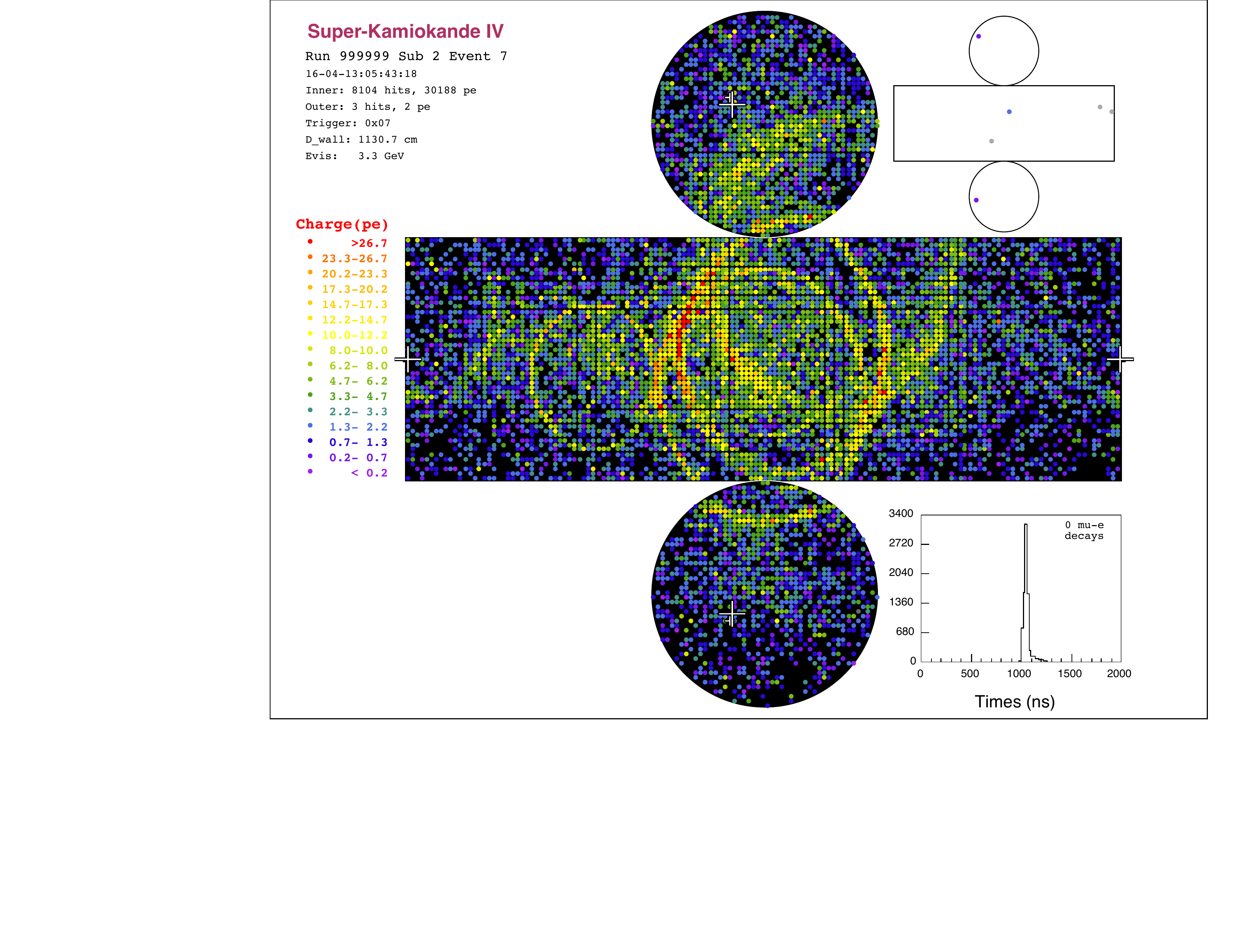}
\caption{Simulation of a single-ring CC background event (top) with 2.8 GeV visible energy in the ID, a multi-ring NC background event (middle) with 2.2 GeV visible energy in the ID, and a CC $\nu_\tau$ event (bottom) with 3.3 GeV visible energy in the ID. The tau signal event produces multiple rings, making it different from the single-ring background event. The background event with multi-rings has a similar pattern to the signal event, and requires more effort to statistically distinguish.}
\label{fig:eventdisplay}
\end{figure}

As described in Section \ref{sec:mc}, tau leptons produced in CC $\nu_{\tau}$ interactions decay quickly to secondary particles. Because of the short lifetime of tau lepton,  it is not possible to directly detect them in Super-K. The decay modes of the tau lepton are classified into leptonic and hadronic decay based on the secondary particles in the decay. The leptonic decays produce neutrinos and an electron or a muon. These events look quite similar to the atmospheric CC $\nu_e$ or $\nu_\mu$ background. The hadronic decays of the tau are dominant and produce one or more pions plus a neutrino. The existence of extra pions in the hadronic decays of tau allows the separation of the CC $\nu_\tau$ signal from CC $\nu_{\mu}$, CC $\nu_{e}$ and NC background. As shown in Fig. \ref{fig:eventdisplay}, CC $\nu_\tau$ events typically produce multiple rings in the  detector. Multiple-ring events are relatively easy to separate from single-ring atmospheric neutrino events. However, the multi-ring background events, resulting from multi-pion/DIS atmospheric neutrino interactions, are difficult to distinguish from the tau signal. Simple selection criteria based on kinematic variables do not identify CC $\nu_\tau$ events efficiently. In order to statistically identify events with the expected characteristics that differentiate signal and background, a multivariate method is applied in this analysis. Specifically, a multi-layer perceptron (MLP) method is used. It is implemented in the ROOT-based TMVA \cite{Hocker:2007ht} library, and was also used in our previously published $\nu_\tau$ search\cite{Abe:2012jj}.

A multilayer perceptron is a feed-forward artificial neural network (NN), which maps between a set of inputs and a set of outputs. It is typically organized in layers of inter-connected neurons with one or more hidden layers between the input and output layer. Neurons in the input layer receive inputs, then normalize the inputs and forward them to the neurons in the first hidden layer. Each layer is fully linked to neurons in the next one with weighted connections. The output of a neuron is scaled by the connecting weights and fed forward to the neurons in the next layer. A MLP has the ability to learn through training, during which the weights in the network are adjusted. Once trained with representative training data, the MLP can be applied to new, unseen data.

\par A MLP is used in this analysis, which has seven inputs, ten neurons in one hidden layer, and one output. It takes seven input variables for both the CC $\nu_\tau$ signal and atmospheric neutrino background to produce a single discriminating output variable that separates signal and background. To prepare the MLP algorithm for event identification, three stages are required. They are referred to as training, testing, and analysis. Separate sets of signal and background MC are used in each stage. We describe the MLP that we implemented for this analysis below.
\par
\begin{figure}[htbp] 
   \centering
   \includegraphics[width=0.4\textwidth]{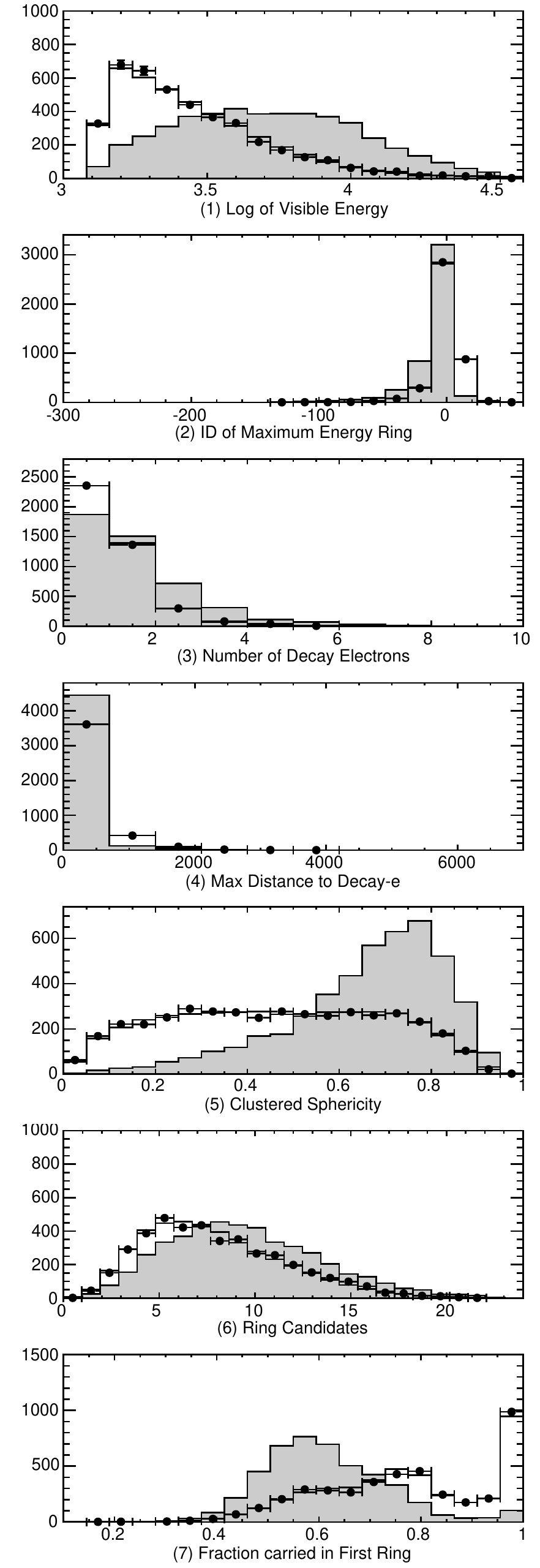} 
   \caption{The input variables to the neural network for the downward MC (black histogram) and data (black dots), along with tau MC (gray shaded histogram) in SK-I through SK-IV. The tau signal is normalized for equal statistics.}
   \label{fig:Downward-variables}
\end{figure}

 Seven variables are used as inputs to the MLP based on the expected separation between signal and background in these variables. The seven variables are
\begin{itemize}
\item[(1)] The $\log_{10}$ of the total visible energy in MeV ($E_{\rm vis}$) of the event. Due to the energy threshold of CC $\nu_\tau$ interactions and the large mass of tau lepton, the signal events are expected to have higher average visible energy than background. The $E_{\rm vis}$ spectrum of CC $\nu_\tau$ signal peaks around 4 GeV, as shown in Fig. \ref{fig:Downward-variables}. By contrast, the $E_{\rm vis}$ spectrum of background falls with increasing $E_{\rm vis}$.

\item[(2)] The particle identification likelihood parameter of the ring with maximum energy. Tau leptons decay quickly to daughter particles after production through leptonic and hadronic decays. Except for the leptonic decay to a muon, most decay channels have at least one showering particle. A showering particle has a negative value in the definition of particle identification likelihood, compared with a positive value for a non-showering particle. The particle identification of the most energetic ring for the signal has a distribution mostly in the negative region, while the background has a broad distribution in both negative and positive regions.

\item[(3)] The number of decay electron candidates in the event. Naively, we expect more decay electrons for signal from pion decays which are produced in hadronic tau decays. This variable does not depend on ring reconstruction, so it is relatively independent of most other variables.

\item[(4)] The maximum distance between the primary interaction point and any decay electron from a pion or muon decay. Energetic muons can travel a long distance in water. Therefore, CC $\nu_\mu$ background involving a high energy neutrino is expected to have a large distance between the primary interaction point and the decay electron from the muon. In comparison, the pions from hadronic tau decay are expected to have smaller momentum, resulting in a smaller value of the variable. 

\item[(5)] The clustered sphericity of the event in the center of mass system. Sphericity is a measure of how spherical an event is. A perfectly isotropic event has sphericity 1, while a perfectly one-directional event has sphericity 0. We follow the definition from \citep{Sjostrand:2006za}, defining the spherical tensor as
\begin{equation}
S^{\alpha\beta} = \frac{\sum_{i}p_{i}^{\alpha}p_{i}^{\beta}}{\sum_{i}p_{i}^{2}},
\end{equation}
where $\alpha$,$\beta= 1, 2, 3$ are three cartesian momentum vectors pointing
to binned photoelectric charge in the event. Sphericity is then constructed by finding the eigenvalues, $\lambda_1 > \lambda_2 > \lambda_3$, of the tensor.
\begin{equation}
S = \frac{3}{2}(\lambda_2 + \lambda_3).
\end{equation}
\par The hadronic decay of the heavy tau lepton is more isotropic than a typical $\nu_\mu$ or $\nu_{e}$ background. The spectrum of sphericity is centered near 1 for signal, while the spectrum for background has an almost flat distribution between 0.1 and 0.8. 
\item[(6)] The number of possible Cherenkov ring fragments. In the ring reconstruction, these ring candidates are formed using a method based on a Hough transformation to find rings. We expect more ring candidates for signal because of the multiple charged particles and pions in hadronic tau decay. This variable is sensitive to even partial ring fragments.
\item[(7)] The fraction of the total number of photoelectrons carried by the most-energetic ring in an event. This variable is calculated from the number of photoelectric charge in each PMT ($q_i$) and the reconstructed vertex and direction of an event as:
\begin{equation}
{\rm rfrac} = \frac{\sum_{\theta_i < 48^\circ}q_i}{\sum q_i},
\label{eq:newrfrac}
\end{equation}
where $\theta_i$ is the angle between the reconstructed direction of the first ring  and the direction of the reconstructed vertex to a PMT. The variable calculates the ratio of charge within 48$^\circ$ of the direction of first ring in the event. The variable is expected to be smaller for the signal because energy is carried by multiple particles in the hadronic decay of the tau.
\end{itemize}

Because the oscillation-induced tau events come from below, the downward sample is expected to have no tau events. Therefore, the data in the downward sample can be used to study the extent to which the atmospheric neutrino simulation for background events agrees with data. Figure \ref{fig:Downward-variables} shows the seven variables for data and MC in the downward sample, along with the expected tau signal. The data and background MC have good agreement.

\par The neural network is trained with a set of signal and background MC with the target of output = 1 for signal and 0 for background. The weights in the MLP are adjusted iteratively during the training such that the difference between the actual output of the MLP and the target is minimized. 

\par
\begin{figure}[!htbp]
\centering
\includegraphics[width=0.45\textwidth]{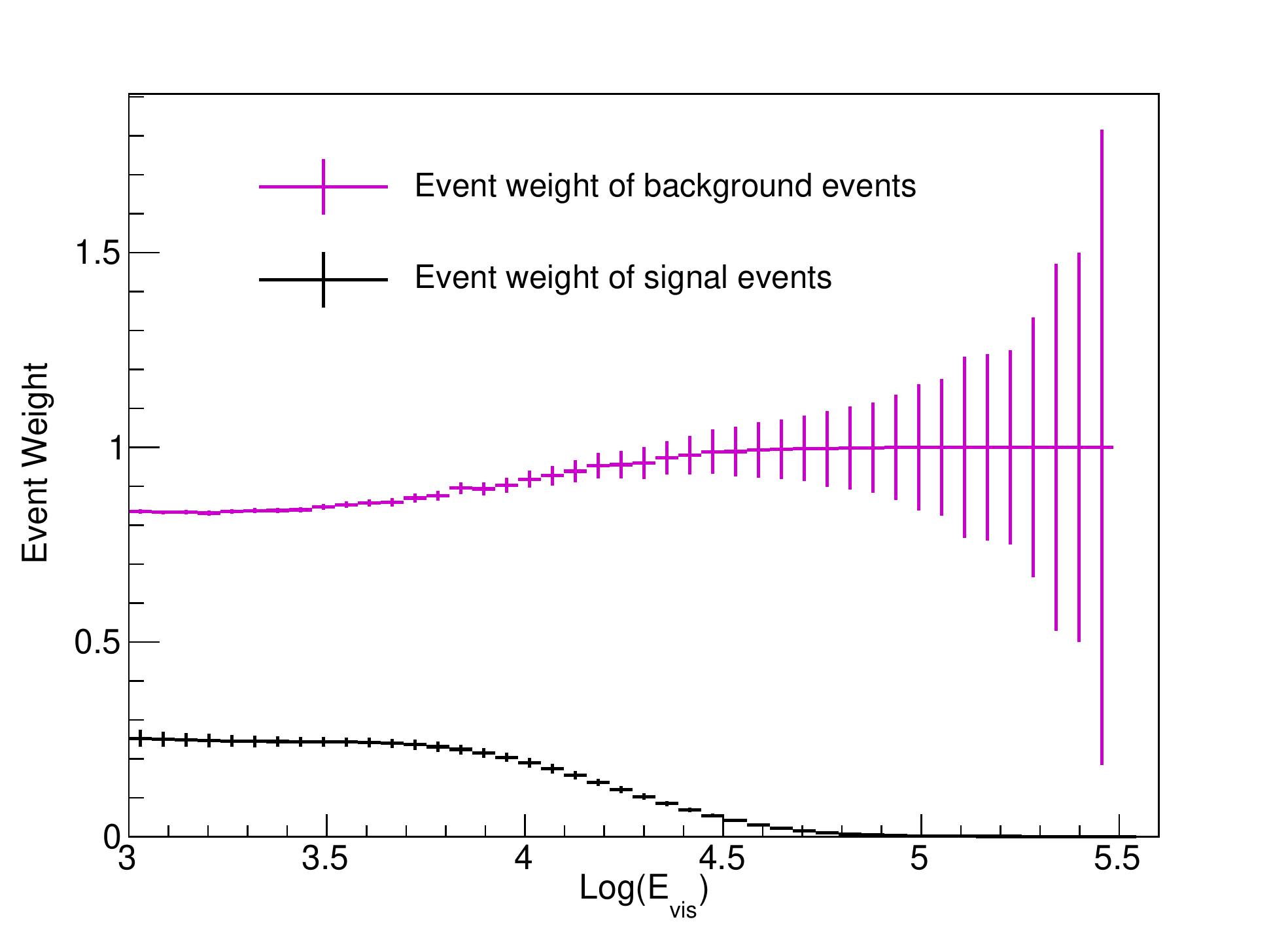}
\caption{Event weights for atmospheric neutrino background (magenta) and CC $\nu_\tau$ signal (black) as a function of $log_{10}(E_{\rm vis})$ in the training sample. The event weight is calculated as an average value of oscillation weights for each bin of $log_{10}(E_{\rm vis})$.}
\label{fig:energyweight}
\end{figure}

 The MC is generated with a Honda flux that calculates the fluxes of neutrinos at production in the atmosphere \citep{Honda:2015fha}. Therefore, the events need to be weighted with oscillation probabilities before being fed to the MLP. Tau neutrinos from oscillations are expected to mostly come from below because the oscillation length of neutrinos in excess of the tau threshold is at least 4,100 km. Used naively, the oscillation weight described in Section \ref{sec:mc} will encode this up-down asymmetry into the signal MC, the training sample is instead weighted based on the average probability at its visible energy, as shown in Fig. \ref{fig:energyweight}. In other words, instead of weighting each MC event with its oscillation probability, the weight is calculated as an average of the oscillation probabilities for MC events in each bin of  $log_{10}(E_{\rm vis}$). In this way, the upward and downward samples are treated the same in the training process, while the overall weight is still correct.  Moreover, since the weights of the downward signal simulation are not set to zero, the whole of signal MC statistics are preserved for training.

\par
\begin{figure}[!ht] 
   \centering
   \includegraphics[width=0.4\textwidth]{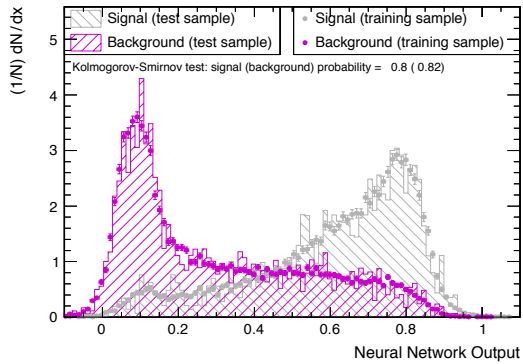} 
   \caption{Neural network output of training (filled histogram) and test (dots) sample for background (magenta) and signal (gray) simulations in SK-IV.}
   \label{fig:NNtrainingandtest}
\end{figure}

 During training, a testing dataset is used as validation to avoid overtraining. Figure \ref{fig:NNtrainingandtest} shows the neural network output for background and signal with the training and testing samples. The clear separation of signal and background in both samples demonstrates that the MLP learned to separate signal from background. Also, the good agreement between training and testing samples shows that it is properly trained.

\par
\begin{figure}[!ht] 
   \centering
   \includegraphics[width=0.4\textwidth]{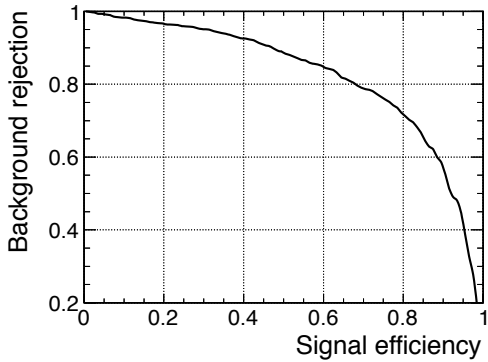} 
   \caption{Efficiencies of signal selection and background rejection by cutting the NN output in SK-IV.}
   \label{fig:sigeffbgref}
\end{figure}

 The testing sample is also used to plot the efficiencies of signal selection and background rejection by cutting on NN output, as shown in Fig. \ref{fig:sigeffbgref}. By varying the cut on NN output, the efficiencies of signal selection and background rejection can be changed. When selecting tau-like events from the events after reduction by requiring the NN output be greater than 0.5, 76$\%$ of signal events and only 28$\%$ of background remain. Table~\ref{tab:interactionmodes} summarizes the breakdown of the interaction modes in different samples, including the fraction for tau and non-tau like samples. Table~\ref{tab:pdgtaudecay} summarizes the decay modes of the largest branching fractions and the fraction of tau-like events in each mode. These efficiencies are only shown to assess the performance of the neural network in selecting tau events. No cut is used in the following analysis.

\begin {table*}[!ht]
\centering
\begin{tabular}{ c@\quad @\quad c@\quad @\quad  c@\quad @\quad c }
\hline
\hline
Interaction Mode   & non-tau-like & tau-like& All\\
  \hline
  CC $\nu_e$& 3071.0 & 1399.2 &  4470.2\\
  CC $\nu_\mu$& 4231.9 & 783.4& 5015.3\\
  CC $\nu_\tau$& 49.1 & 136.1 & 185.2\\
  NC & 291.8& 548.3 & 840.1 \\
  \hline
  \hline
\end{tabular}
\caption{The break down of interaction modes of both background and expected signal shown in number of events in simulation scaled to SK-I through SK-IV live time. By cutting the NN output at 0.5, each mode is separated into tau-like (NN$>$0.5) and non-tau-like (NN$<$0.5).}
\label{tab:interactionmodes}
\end{table*}

\begin {table*}[!htbp]
\centering
\begin{tabular}{ l @\quad c@\quad c@\quad c}
\hline
\hline
  Decay mode & Branching ratio ($\%$) &  Tau-like fraction ($\%$) & Branching ratio$\times$Tau-like fraction ($\%$)\\
  \hline
  $e^{-}\bar{\nu}_{e}\nu_{\tau}$ & 17.83 & 67.3$\pm$2.2& 12.0$\pm$0.4 \\
  $\mu^{-}\bar{\nu}_{\mu}\nu_{\tau}$ & 17.41 & 42.6$\pm$2.6 & 7.2$\pm$0.5\\
  $\pi^{-}\nu_{\tau}$ & 10.83 & 84.7$\pm$3.8  & 9.2$\pm$0.4\\
  $\pi^-\pi^0\nu_{\tau}$ & 25.52 & 81.0$\pm$2.1&  20.7$\pm$0.5\\
  $3\pi\nu_{\tau}$ & 18.29 & 88.7$\pm$2.5&  16.2$\pm$0.5\\
  others & 10.12 & 90.5$\pm$3.4& 9.2$\pm$0.3\\
  \hline
  \hline
\end{tabular}
\caption{Decay modes of tau leptons with branching ratio adapted from \citep{Olive:2016xmw}, along with the fraction of tau-like events and the product of branching ratio and tau-like ratio in each mode in the Super-K simulation.}
\label{tab:pdgtaudecay}
\end{table*}

\par The analysis sample is finally processed with the trained MLP. Unlike the training and testing processes, no information is given to the MLP regarding the composition of the analysis samples as either signal or background. The analysis sample processed with the trained MLP is used in this analysis.
\par Table \ref{tab:MCsamplesize} summarizes the quantities of signal and background MC samples used for each stage. For each SK run period, separate MLPs are trained, tested and analyzed. Real data in each SK run period are processed with the corresponding trained MLP.

\begin {table}[!ht]
\centering
\begin{tabular}{ c@\quad c@\quad c@\quad c }
\hline
\hline
   &training &testing & analysis  \\
  \hline
  Signal & $\sim$25,000& 1,500&  $\sim$6,600 (100 yr.)\\
  Background & $\sim$32,000&1,500 & $\sim$82,000 (100 yr.)\\
  \hline
  \hline
\end{tabular}
\caption{Monte Carlo sample sizes for each stage of the MLP. The same sample sizes are used for all SK running periods (I through IV).}
\label{tab:MCsamplesize}
\end{table}

\section{Analysis}\label{sec:analysis}
\subsection{Search for atmospheric tau neutrino appearance}\label{sec:tauapp}
\par
\begin{figure}[!ht]
\centering
\includegraphics[width=0.45\textwidth]{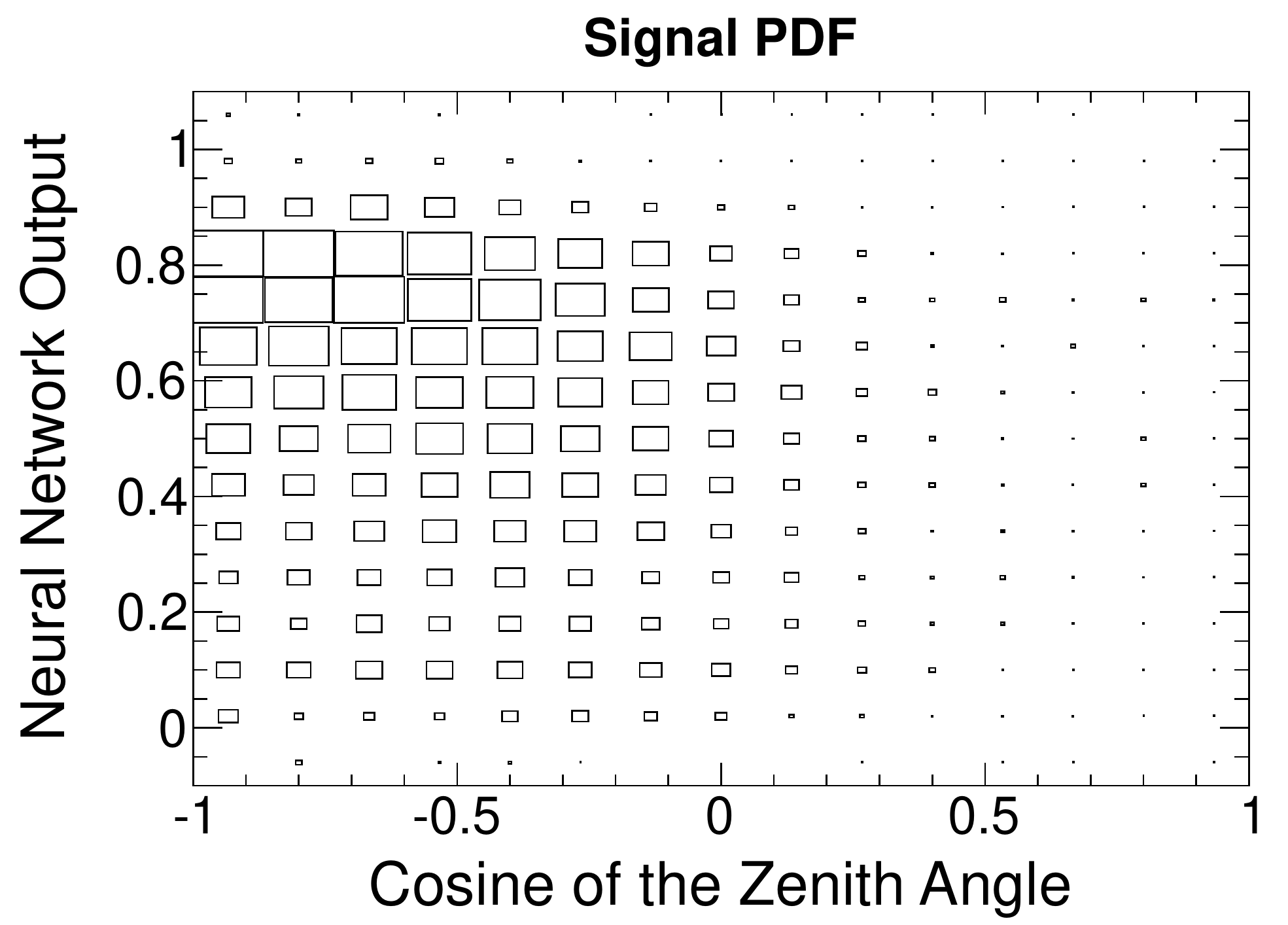}
\includegraphics[width=0.45\textwidth]{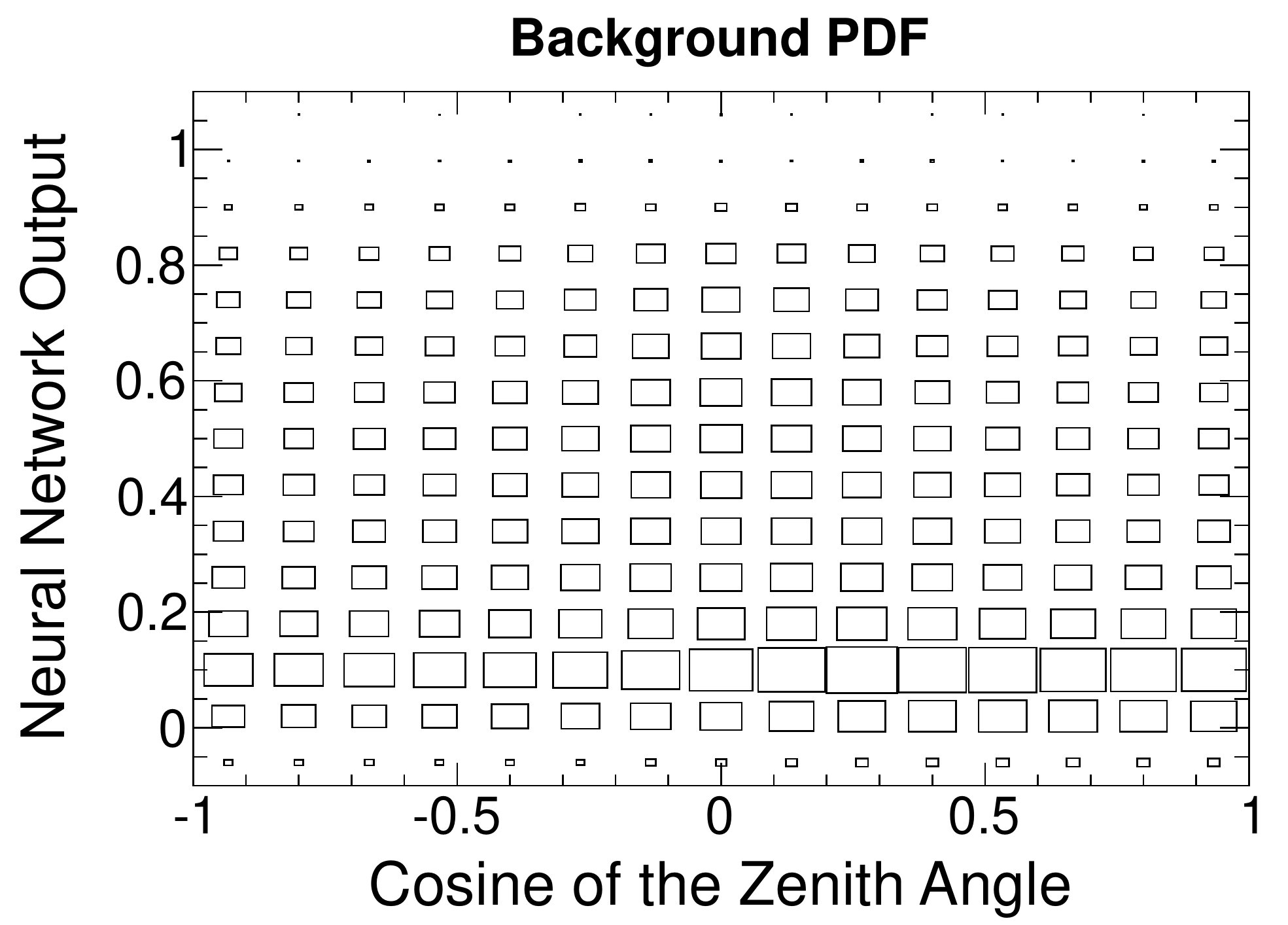}
\caption{Two-dimensional probability distribution likelihood as a function of zenith angle and neural network output for tau (top) and background (bottom) built with SK-IV MC.}
\label{fig:2dpdf}
\end{figure}

\begin{figure}[htbp] 
   \centering
   \includegraphics[width=0.45\textwidth]{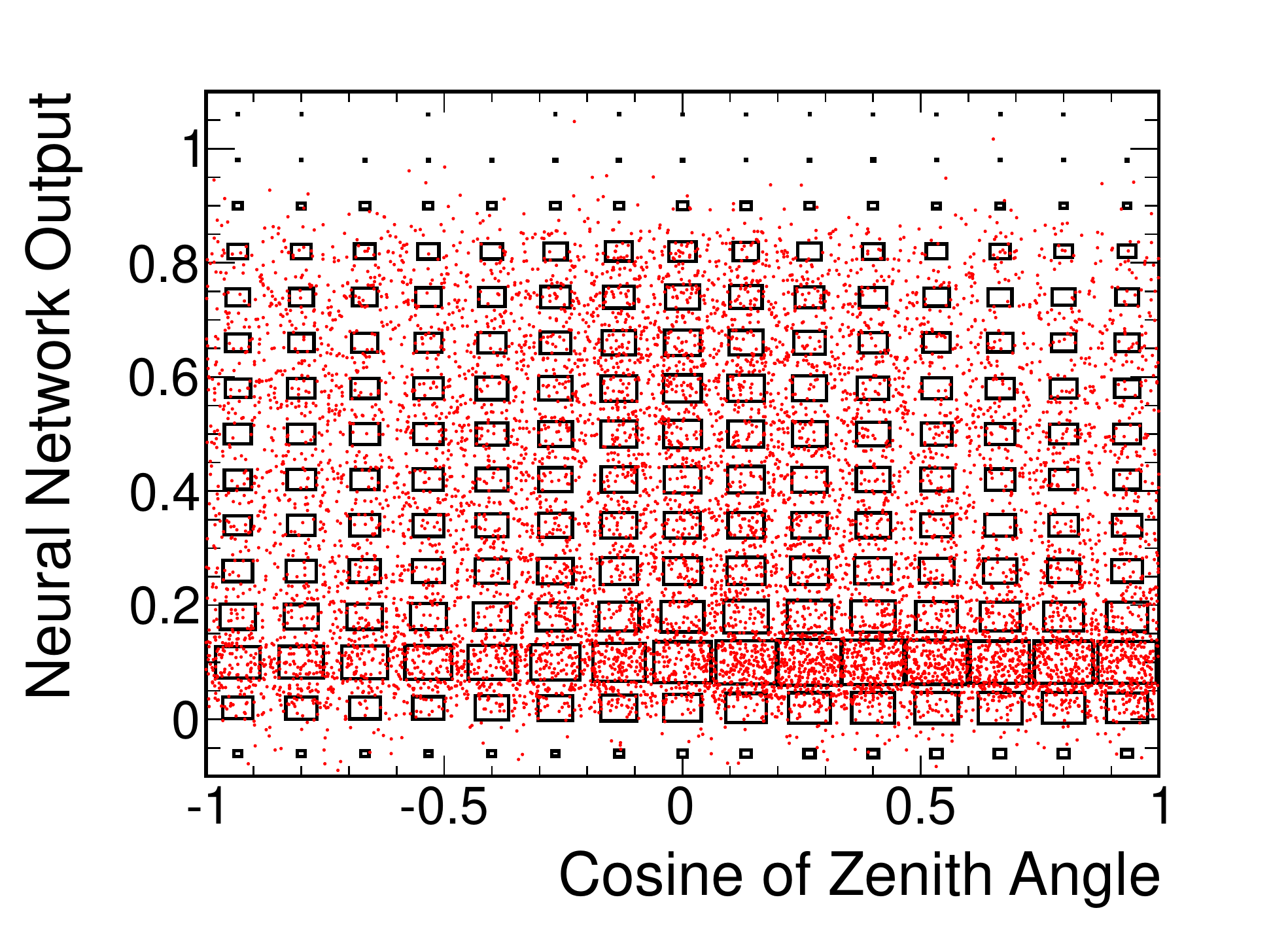} 
   \caption{A combination of signal PDFs and background PDFs with both tau normalization and background normalization equal to 1, overlaid with every individual event shown in red dot for SK-I to SK-IV data.}
   \label{fig:datapoints}
\end{figure}

 To search for atmospheric tau neutrino appearance, the data is fit to a combination of the expected tau signal resulting from neutrino oscillations and atmospheric neutrino background with neutrino oscillations. In order to extract maximum information from the sample, the analysis uses a two-dimensional unbinned maximum likelihood fit implemented in RooFit \cite{Verkerke:2003ir}. Using two-dimensional histograms of the neural network output and the
reconstructed zenith angle of the events, two-dimensional probability distribution functions (PDFs) are built for background and tau signal. The probability density follows the normalized bin contents in the histograms. Figure \ref{fig:2dpdf} is an example of the 2D distributions for oscillated signal on the top and background on the bottom. The horizontal axis of the plots is the cosine of the reconstructed zenith angle as determined by the energy-weighted sum of the ring directions in the event. The vertical axis is the NN output, in which tau-like events have a
value close to 1 and non-tau-like events have a value close to 0. The signal events (top panel) are primarily tau-like and come from below (cosine of the zenith angle, $\Theta$, less than zero), while the background (bottom panel) is more non-tau-like and come from all directions. The amount of signal and background events can be adjusted by varying the normalization of the distributions. Figure~\ref{fig:datapoints} shows a combination of signal PDFs and background PDFs for SK-I to SK-IV with both tau normalization and background normalization equal to 1, with the data overlaid on the combined PDF.


\par
\begin{figure}[htbp]
   \centering
   \includegraphics[width=0.45\textwidth]{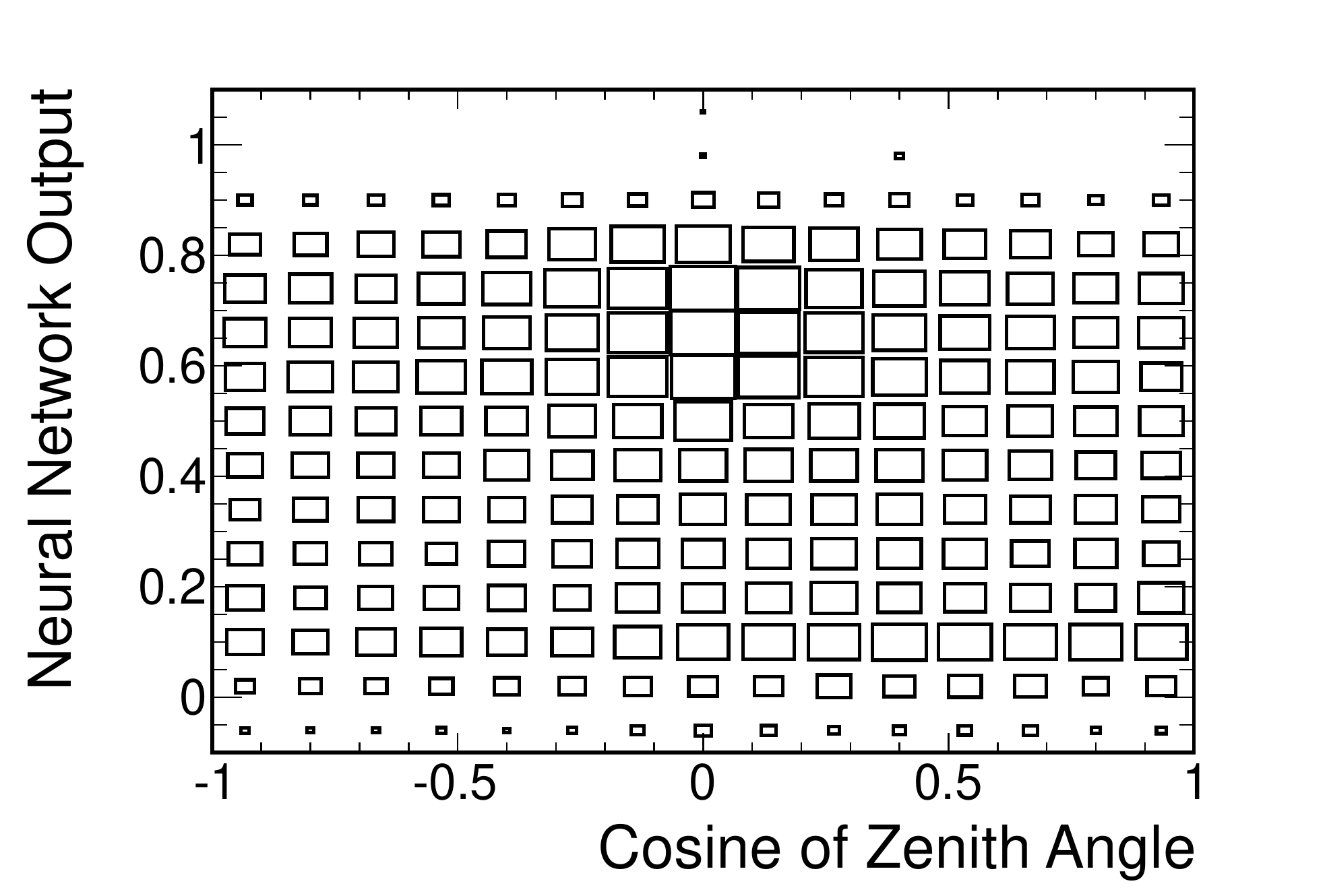} 
   \caption{An example histogram of the PDF of DIS cross section uncertainty for SK-IV background. The vertical axis is the output of the NN, the horizontal axis is the cosine of the zenith angle distribution.}
   \label{fig:sysexample}
\end{figure}

 The systematic errors used in this analysis are selected from the systematic errors in the Super-K atmospheric neutrino analysis \cite{Abe:2012jj}. Only systematic errors for which the maximum bin content change in the two-dimensional signal/background histograms after shifting that systematic error by 1$\sigma$ is larger than 2.5$\%$ are considered. After the re-evaluation, 28 systematic errors from atmospheric neutrino analysis are considered in the fit. The systematic errors are summarized in Table \ref{tab:newfitsyscommon} and \ref{tab:newfitsysruns}. In order to simultaneously fit the systematic errors, a set of PDFs are built for each systematic error with the same structure as the PDFs for signal and background in Fig \ref{fig:2dpdf}. The construction of PDFs for systematic errors is based on the three-flavor oscillation framework in Ref. \cite{Wendell:2010md}. The framework is capable of estimating the resulting change in a given event distribution after changing in a single systematic error. A two-dimensional histogram of NN output and reconstructed zenith angle is built which results from a 1$\sigma$ change in each systematic error. An example of a two-dimensional distribution for a 1$\sigma$ change is shown in Fig. \ref{fig:sysexample}. The size of a systematic error determines the normalization of the distribution, which adjusts the size of the systematic error in the fit. 

\par Since the uncertainties in oscillation parameters can also change the measured results and significance, $\sin^2 2\theta_{23}$ and $\Delta m_{32}^2$ are also varied within the 1$\sigma$ limits of Super-K atmospheric neutrino oscillation analysis result \cite{Abe:2012jj}. The value of $\Delta m^2_{32}$ is varied between 1.92$\times$10$^{-3}$ and 2.22$\times$10$^{-3}$ eV$^2$, and $\sin^22\theta_{23}$ is varied between 0.93 and 1.0. The value of $\sin^2(2\theta_{13})$ is varied within 1$\sigma$ limit of the combined Daya Bay \cite{An:2012eh} and RENO \cite{Ahn:2012nd} measurements of $\sin^22\theta_{13}$=0.099$\pm$0.014. For this analysis, the value of $\delta_{CP}$ is set to be zero, varying its value from 0 to $2\pi$ results in less than 1\% change in the number of fitted tau events. The analysis is performed for both normal and inverted hierarchy. The data is fitted to the sum of background PDF, signal PDF and systematic PDFs varying the normalizations using a RooFit-based \cite{Verkerke:2003ir} unbinned likelihood fit algorithm as
\begin{equation}
Data = PDF_{BG} + \alpha \times PDF_{tau} + \sum \epsilon_{i}\times PDF_{i},
\end{equation}
where $\alpha$ is the normalization of the tau signal, with 0 meaning no-tau-appearance and 1 meaning the expected tau appearance based on the neutrino oscillation parameters assumed in the simulation. The size of the i$^{th}$ systematic error in the fit, $\epsilon_i$, has a Gaussian univariate constraint in the fit. The PDFs of systematic errors are built separately for signal and background, but are combined together in the fit because the normalizations of both PDFs are adjusted by the same normalization factor $\epsilon_i$ simultaneously.
\begin{table*}
\centering
\caption{Systematic errors used in the tau neutrino appearance search that are common to all Super-K run periods. The systematic errors are ordered by the maximum fractional change in the bins of the two-dimensional event distribution after shifting the systematic error by 1$\sigma$, and only systematic errors with the maximum fractional change larger than 2.5$\%$ are shown. The estimated 1-$\sigma$ error size  is shown in percentage.}
\begin{tabular}{l@{\hskip 1in} c }
\hline
\hline
Systematic error& $\sigma$ ($\%$)\\
\hline
NC/CC ratio & 20\\
DIS $q^2$ dependence for low W & 10 \\
Meson exchange current& 10  \\
1$\pi$ axial coupling & 10 \\
DIS $q^2$ dependence for high W & 10 \\
Coherent $\pi$ cross section & 100\\
Flux normalization ($E_\nu>$ 1GeV) & 15 \\
1$\pi$ background scale factor & 10 \\
1$\pi$ axial form factor& 10\\
CCQE cross section & 10 \\
Single pion $\pi^0/\pi^{\pm}$ ratio & 40\\
$\bar{\nu}_{\mu}/\nu_\mu$ ratio ($E_\nu > $ 10 GeV) & 15 \\
$\nu/\bar\nu$ ratio ($E_{\nu}>$ 10 GeV) & 5 \\
DIS cross section (E$_{\nu}<$ 10 GeV)& 10  \\
FC multi-GeV normalization & 5 \\
$\bar\nu_e/\nu_e$ ratio ($E_{\nu}> 10 $GeV) & 8 \\
$K/\pi$ ratio & 10 \\
Single meson cross section & 20 \\
Single-pion $\bar\nu/\nu$ ratio & 10 \\
Horizontal/vertical ratio & 1 \\
CCQE $\nu/\bar\nu$ ratio & 10 \\
DIS cross section & 5 \\
Matter effect & 6.8  \\
Neutrino path length& 10 \\
\hline
\hline
\end{tabular}
\label{tab:newfitsyscommon}
\end{table*}

\begin{table*}
\centering

\caption{Systematic errors used in the tau neutrino appearance search that are dependent on Super-K run periods. The systematic errors are ordered by the maximum fractional change in the bins of the two-dimensional event distribution after shifting the systematic error by 1$\sigma$, and only systematic errors with the maximum fractional change larger than 2.5$\%$ are shown. The estimated 1-$\sigma$ error size  is shown in percentage.}
\begin{adjustbox}{max width=\textwidth}
\begin{tabular}{l@{\hskip 0.4in} c@{\hskip 0.4in} c @{\hskip 0.4in}  c @{\hskip 0.4in} c @{\hskip 0.4in} c }
\hline
\hline
    & & SK-I & SK-II & SK-III & SK-IV\\
    & & $\sigma$ ($\%$)&  $\sigma$ ($\%$)&  $\sigma$ ($\%$)& $\sigma$ ($\%$)\\
\hline
Multi-ring e-like background & & 12.1 & 11.1 & 11.4 & 11.6 \\
Multi-ring PID  & Multi-GeV $e$-like & -2.9 & -3.9 & 2.7  & -1.6\\
                & multi-GeV $\mu$-like & 6.5 &9.7 & -4.9 & 3.3\\
1-ring e-like background & & 13.2& 38.1 & 26.7 & 17.6 \\
Ring separation & Multi-GeV $e$-like & 3.7 & 2.6 & 1.3 & 1.0\\
                & Multi-GeV $\mu$-like & 1.7 & 1.7& 1.0 & -1.2\\
                & Multiring Multi-GeV $e$-like & -3.1 & -1.9& -1.1 & 0.9\\
                & Multiring Multi-GeV $\mu$-like & -4.1 & -0.8 & -2.1 & 2.4\\

\hline
\hline
\end{tabular}
\end{adjustbox}
\label{tab:newfitsysruns}
\end{table*}

\par The fit is performed jointly with all data periods being fit at the  same time. First, we present the results of the fit assuming the normal hierarchy of neutrino mass splitting. Relative to the expectation of unity, the tau normalization is found to be 1.47$\pm$ 0.32 (stat+syst) in the joint fit. This corresponds to a significance level of 4.6$\sigma$ of rejection the hypothesis of no-tau-appearance. To estimate the statistical uncertainty of the fitted tau normalization, the systematic errors are excluded from the fit, and the tau normalization is found to be 1.41$\pm$0.28. Therefore, the total uncertainty is dominated by the statistical uncertainty. The measured significance is larger than the expected significance of 3.3$\sigma$ because more events are measured than expected. The number of tau events observed is evaluated
after the fit by adding the tau events in the signal PDF rescaled by the fitted tau normalization and tau events in the systematic PDFs rescaled by the fitted values of systematic errors. The number of tau events is found to be 290.8 in the sample selected for this analysis. After correcting for efficiency, the observed number of fitted CC $\nu_\tau$ events over the entire running periods is calculated to be 338.1$\pm$72.7 (stat+sys), compared with an expectation of 224.5$\pm$57.2 (sys) interactions.
\par The fit is repeated with the inverted hierarchy when calculating the oscillation probabilities, resulting in a reduction in the expected number of $\theta_{13}$ induced upward-going electron neutrino. Under the assumption of inverted hierarchy, the fit results in a higher fitted value of tau normalization, 1.57$\pm$0.31 and a correspondingly higher significance of 5.0$\sigma$. The higher fitted tau normalization is due to the reduction in $\theta_{13}$-induced upward-going electron neutrinos when calculating the oscillation probabilities under the assumption of the inverted hierarchy.
\par
\begin{figure*}[htbp]
   \centering
   \includegraphics[width=0.8\textwidth]{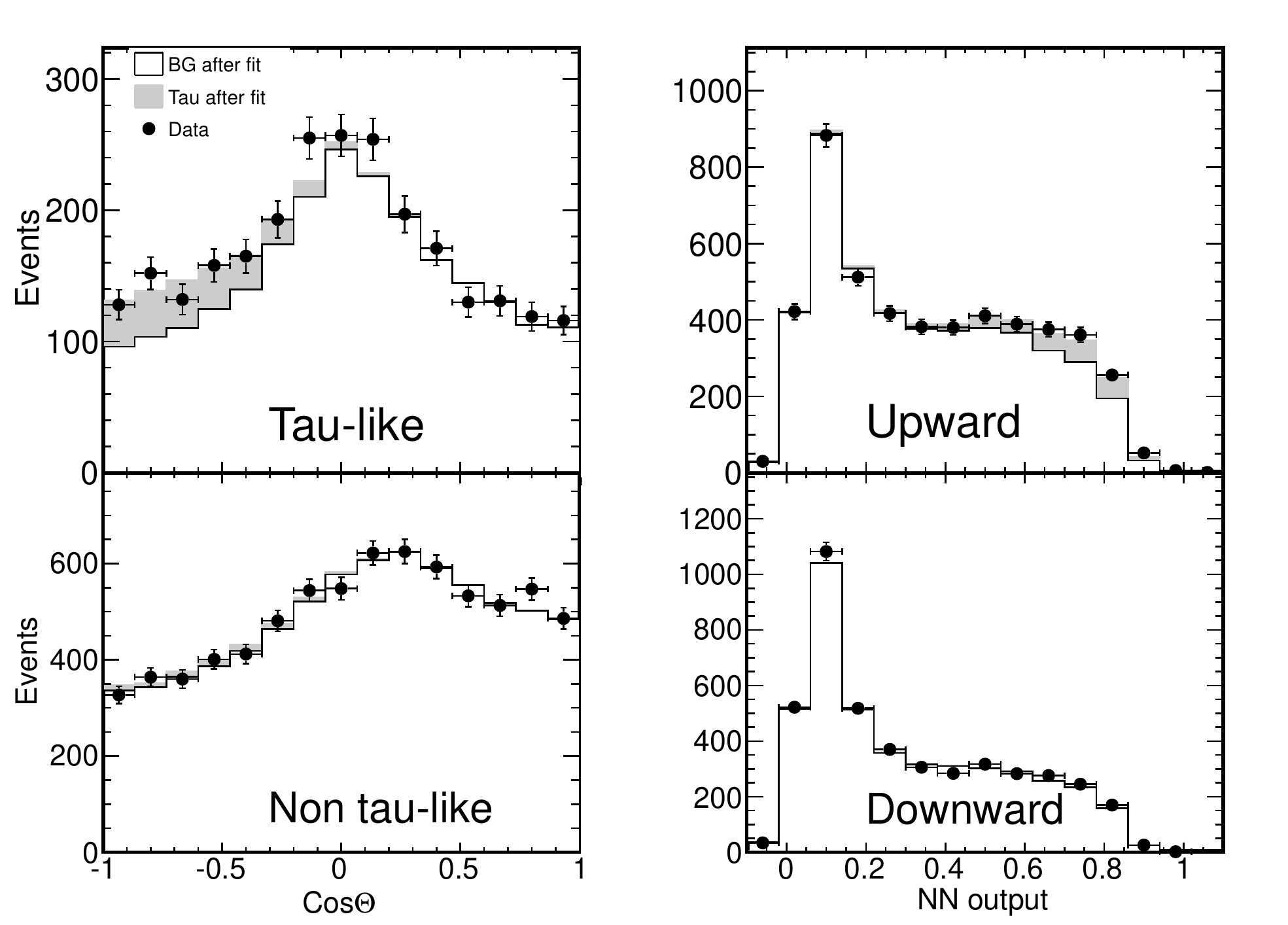} 
   \caption{Fit results, assuming the normal hierarchy, showing binned projections in the NN output and zenith angle distribution for tau-like ($\textrm{NN}>0.5$), upward-going [$\cos\Theta<-0.2$], non-tau-like ($\textrm{NN}<0.5$) and downward-going [$\cos\Theta>0.2$] events for both the two-dimensional PDFs and data. The PDFs and data sets have been combined from SK-I through SK-IV. The fitted tau signal is shown in gray.}
   \label{fig:fourpanelnew}
\end{figure*}

 The results of the final combined fit are examined graphically by plotting the binned projections of the fitted results. Figure \ref{fig:fourpanelnew} demonstrates the projection in zenith angle for both tau-like (NN $\textrm{output}>0.5$) and non-tau-like (NN $\textrm{output}<0.5$) events, along with the projections in NN output for both upward-going ($\cos\Theta<-0.2$) and downward-going ($\cos\Theta>0.2$) events. In these plots, the signal PDFs have been rescaled to the fitted normalization values, and PDFs of systematic errors for signal and background have been rescaled by the fitted magnitudes of systematic errors and added to signal and background respectively. The fitted tau signal is shown in gray. All distributions have good agreement between data and MC simulations. In these plots, the PDFs and data from all of the run periods are combined.

\subsection{Charged-current tau neutrino cross section measurement}
\label{sec:xsec}
This sample of CC $\nu_\tau$ interactions observed in Super-K offers the opportunity to measure the CC $\nu_\tau$ cross section. By scaling the theoretical cross section in the MC simulations to match the data, we can measure the inclusive charged-current tau neutrino cross section in water:
\begin{equation}
\sigma_{measured}= S_{\tau}\times\langle\sigma_{theory}\rangle,
\label{eq:xseceq}
\end{equation}
where $S_\tau$ is the factor that is used to scale the theoretical cross section to match simulations and data. For this analysis, $S_\tau$ is the tau normalization measured in the search for tau neutrino appearance in Section \ref{sec:tauapp}. Therefore, the measured CC $\nu_\tau$ cross section is expressed as:
\begin{equation}
\sigma_{measured} =(1.47\pm0.32)\times\langle\sigma_{theory}\rangle,
\label{eq:2dmeasurementeq}
\end{equation}
$\langle\sigma_{theory}\rangle$ is the flux-averaged theoretical charged-current tau neutrino cross sections used in the NEUT code.
\par
\begin{figure}[!htp] 
   \centering
   \includegraphics[width=0.45\textwidth]{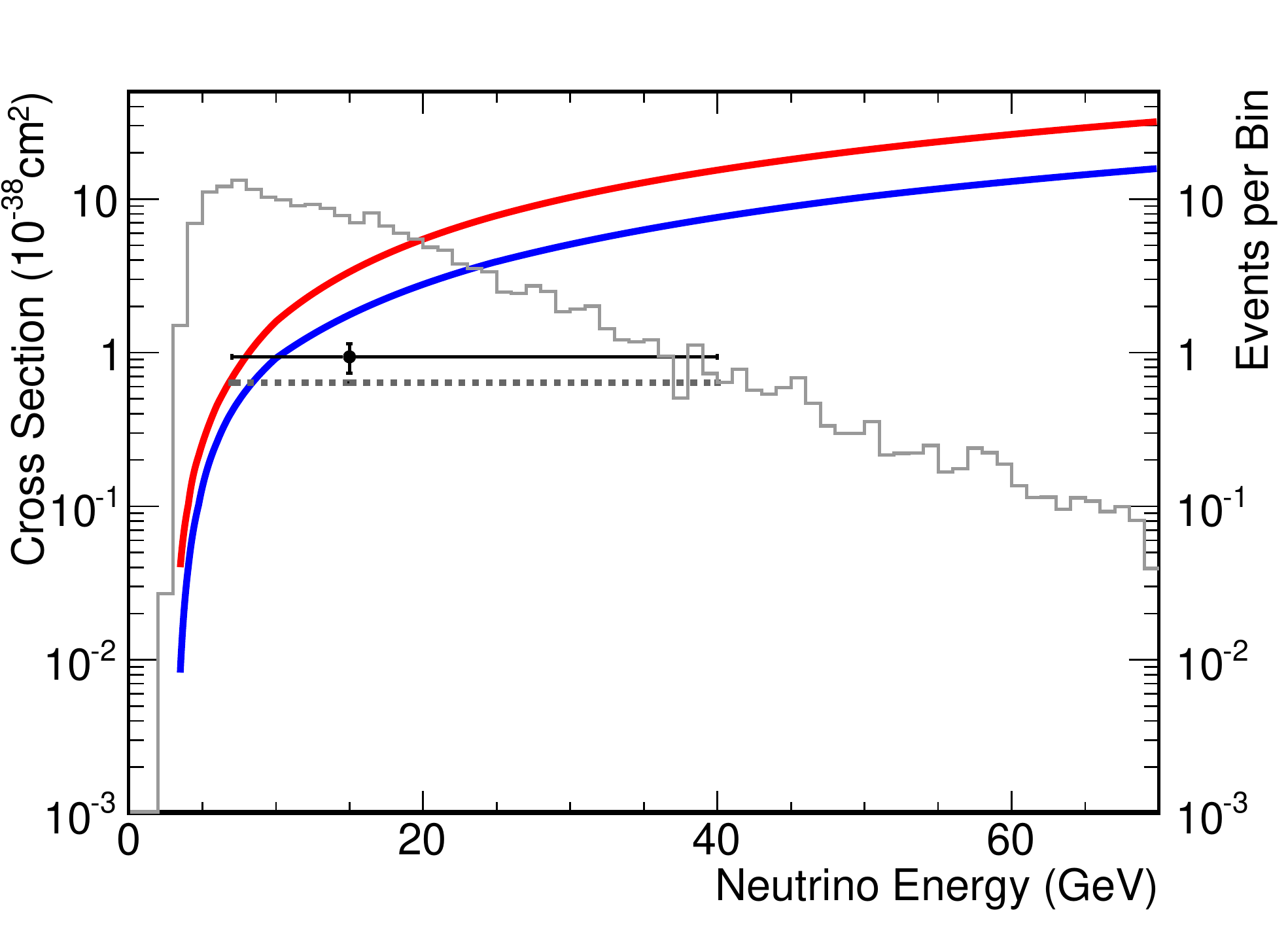} 
   \caption{Measured flux-averaged charged-current tau neutrino cross section (black), together with theoretical differential cross sections ($\nu_\tau$ in red and $\bar\nu_\tau$ in blue), flux-averaged theoretical cross section (dashed gray) and tau events after selection in MC simulations (gray histogram). The horizontal bar of the measurement point shows the 90$\%$ range of tau neutrino energies in the simulation. }
   \label{fig:xsec2Dfit}
\end{figure}

 To calculate the flux-averaged theoretical cross section, the differential CC $\nu_\tau$ cross section as a function of neutrino energy is weighted with the energy spectrum of atmospheric tau neutrinos from neutrino oscillations. Because CC $\nu_\tau$ interactions are not distinguishable from CC $\bar\nu_\tau$ interactions in Super-K, the theoretical cross section is a flux average of $\nu_\tau$ and $\bar\nu_
\tau$ cross sections. The flux-averaged theoretical cross section, $\langle\sigma_{theory}\rangle$, is calculated as:
\begin{equation}
\langle\sigma_{theory}\rangle=\frac{\sum_{\nu_{\tau},\bar{\nu}_{\tau}}\int\frac{d\Phi(E_{\nu})}{dE_{\nu}}\sigma(E_{\nu})dE_{\nu}}{\sum_{\nu_{\tau},\bar{\nu}_{\tau}}\int\frac{d\Phi(E_{\nu})}{dE_{\nu}}dE_{\nu}},
\label{eq:averagexsec}
\end{equation}
where  $\frac{d\Phi(E_{\nu})}{dE_{\nu}}$ is the differential flux of tau neutrinos as a function of neutrino energy as shown in Fig. \ref{fig:taufluxosc}, and $\sigma(E_{\nu})$ is the differential charged-current tau neutrino cross sections used in NEUT code as seen in Fig.~\ref{fig:neuttauxsec}. The range of the integral is determined to be between 3.5 GeV and 70 GeV from the tau neutrino energies in the simulation. As shown in Fig. \ref{fig:xsec2Dfit}, the neutrinos have energies more than 3.5 GeV in the CC $\nu_\tau$ interactions because of the energy threshold, and the expectation of CC $\nu_\tau$ interactions with more than 70 GeV is less than one in the entire run period.

The flux-averaged theoretical charged-current tau neutrino cross
section is calculated to be 0.64$\times 10^{-38}$ cm$^{2}$ between 3.5
GeV and 70 GeV, and thus the measured flux-averaged charged current
tau neutrino cross section:
\begin{equation}
(0.94\pm0.20) \times 10^{-38}{~\rm cm^{2}.}
\label{our_result} 
\end{equation}
\noindent The measured cross section is shown together with
the theoretical cross sections and the MC simulations in
Fig.~\ref{fig:xsec2Dfit}. The measured and theoretical cross section
values are consistent at the 1.5$\sigma$ level.

\subsection{Comparisons of charged-current tau neutrino cross 
section measurement with previously reported results}

Because of the difficulties in tau neutrino production and detection,
charged-current tau neutrino cross sections have not been well
measured. DONUT\cite{Kodama:2000mp} and OPERA\cite{Agafonova:2015jxn}
are the only two experiments that have directly observed
charged-current tau neutrino interactions, and DONUT is the only
experiment that reported a measurement of the cross section. The DONUT
measurement was based on 9 observed charged-current tau neutrino
events with an estimated background of 1.5 events. In DONUT, 800 GeV
protons from the Fermilab Tevatron were used to produce neutrino beam
by colliding with a beam dump, and tau neutrinos were produced via
decays of charm mesons. The mean energy of the detected tau neutrino
interactions was estimated to be 111 GeV, an energy at which deep
inelastic interactions are dominant. Assuming that the DIS
charged-current tau neutrino cross section had a linear dependence on
neutrino energy, DONUT measured the energy-independent slope of the
cross section, $\sigma_{const}$, after correcting for the kinematic
effect of tau lepton mass from the standard model calculation:
\begin{equation}\label{eq:donutxsec}
\sigma(E)= \sigma_{const}\cdot E\cdot K(E),
\end{equation}
where $\sigma(E)$ is the charged-current cross section per nucleon as
a function of neutrino energy, $\sigma_{const}$ is the asymptotic
slope which is constant in $\sigma/E$ for deep inelastic scattering,
and $K(E)$ is the kinematic effect of tau lepton mass. DONUT measured
$\sigma_{const}$ to be $(0.39\pm0.13\pm0.13)\times10^{-38}{\rm
  ~cm^{2}~GeV^{-1}}$ in their final results
paper\cite{Kodama:2007aa}. DONUT was incapable of distinguishing the
charge of the $\tau$ lepton, therefore, the measurement is an average
of the $\nu_\tau$ and $\bar\nu_{\tau}$ cross sections assuming equal
number of $\nu_\tau$ and $\bar\nu_\tau$ in the neutrino flux.

\begin{figure}[h] 
   \centering
   \includegraphics[width=0.45\textwidth]{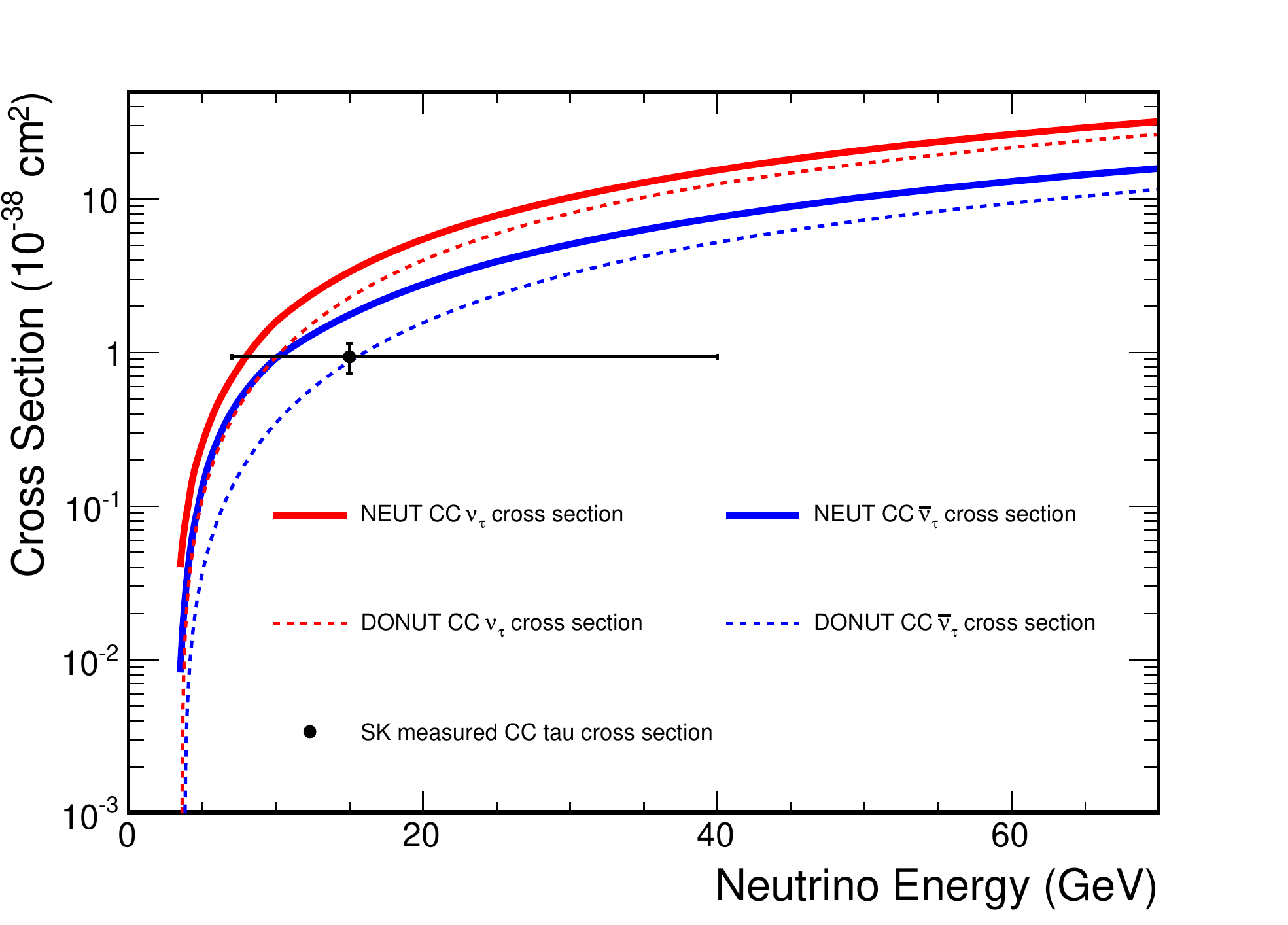} 
   \caption{Comparison of the Super-K measured (marker with error
     bars) and expected tau neutrino cross sections (solid lines) with
     $\sigma(E)$ inferred from DONUT (dashed lines). The DONUT cross
     section considers only DIS, and is is digitized
     from\cite{Kodama:2007aa}.}
   \label{fig:xsecdonut}
\end{figure}

We wish to compare the $\nu_\tau$ cross section measured with
atmospheric neutrinos by Super-K at relatively low energies to that
measured by DONUT with a neutrino beam at higher energies. We
recalculate the DONUT value of $\sigma(E)$ from
Eqn.~\ref{eq:donutxsec} with the kinematic correction $K(E)$
integrated over neutrino energies between 3.5 GeV and 70 GeV and
weighted to the world average ratio of cross sections between
$\nu_\mu$ and $\bar\nu_\mu$\cite{Olive:2016xmw}. The calculated DONUT
value of $\sigma(E)$ is then further weighted by the predicted
$\nu_\tau$ and $\bar\nu_\tau$ flux ratio of 1.11 for atmospheric
neutrino tau appearance at Super-K. The resulting $\sigma(E)$ is shown
in Fig.~\ref{fig:xsecdonut}. The charged-current tau neutrino DIS
cross section inferred from the DONUT published number and rweightd to
lower energy is $(0.37\pm 0.18)\times 10^{-38}{\rm ~cm^{2}}$. This is
smaller than our measurement of $(0.94\pm0.20)\times 10^{-38}{\rm
  ~cm^{2}}$, but the measurements are not yet directly
comparable. DONUT measured the cross section with a neutrino beam that
had a much higher average energy than that of the tau neutrinos in the
atmospheric neutrino flux at Super-K. Quasi-elastic scattering and
resonant pion production is a small component of the DONUT measurement
and was neglected in their calculations. However, the tau neutrino
flux at Super-K has a large component of neutrinos below 10 GeV, where
CCQE and resonant pion production makes a significant contribution to
the detected event rate. We complete the comparison using the
predicted CC DIS fraction in the Super-K sample. According to our
Monte Carlo simulation, the fraction of DIS events in Super-K CC tau
neutrino sample is estimated to be 41$\%$. Therefore, the $\nu_{\tau}$
DIS-only cross section determined by Super-K atmospheric neutrinos is
found to be $(0.40\pm0.08)\times 10^{-38}$ cm$^{2}$ by scaling the
measured cross section in Eqn.~\ref{our_result} by 41\%. This
resulting DIS-only cross-section is comparable and consistent with the
DONUT measurement of the DIS $\nu_{\tau}$ cross section extrapolated
to lower neutrino energy.

\section{Conclusion}\label{sec:conclusion}
\par Using 5,326 days of atmospheric neutrino data in SK-I through SK-IV, Super-K measured the tau normalization to be $1.47\pm0.32$, excluding the hypothesis of no-tau-appearance with a significance of 4.6$\sigma$. A flux-averaged charged current tau neutrino cross section is measured to be  $(0.94\pm0.20)\times 10^{-38}$ cm$^{2}$ for neutrino energy between 3.5 GeV and 70 GeV in Super-K, to be compared with the flux-averaged theoretical cross section of $0.64\times 10^{-38}$ cm$^{2}$. 
Our result is consistent with the previous DONUT result, and is consistent with the Standard Model prediction to within 1.5$\sigma$.

\begin{acknowledgments}
We gratefully acknowledge the cooperation of the Kamioka Mining and Smelting Company. The Super-Kamiokande experiment has been built and operated from funding by the Japanese Ministry of Education, Culture, Sports, Science and Technology, the U.S. Department of Energy, and the U.S. National Science Foundation. Some of us have been supported by funds from the National Research Foundation of Korea NRF-2009-0083526 (KNRC) funded by the Ministry of Science, ICT, and Future Planning, the European Union H2020 RISE-GA641540-SKPLUS, the Japan Society for the Promotion of Science, the National Natural Science Foundation of China under Grants No. 11235006, the National Science and Engineering Research Council (NSERC) of Canada, the Scinet and Westgrid consortia of Compute Canada, and the National Science Centre, Poland (2015/17/N/ST2/04064, 2015/18/E/ST2/00758).
\end{acknowledgments}

\bibliography{bibliography}

\end{document}